# Atomic bonding states of metal and semiconductor elements

Liangjing Ge, Maolin Bo*

Key Laboratory of Extraordinary Bond Engineering and Advanced Materials Technology (EBEAM) of Chongqing, Yangtze Normal University, Chongqing 408100, China

*Corresponding Author: E-mail: bmlwd@yznu.edu.cn (Maolin Bo)

**Abstract**

In this paper, we use density functional theory (DFT) to calculate the deformation electron density of 46 metal and semiconductor elements. The binding-energy and bond-charge model (BBC) model is combined with the tight-binding and density-functional–tight-binding approaches to obtain quantitative information about atomic bonding at the atomic scale and to understand the contributions and effects of deformation energy density, energy shifts, and atomic bonding on the Hamiltonian.

# 1. Introduction

The energies associated with the relaxation, polarization, and localization of electron wave functions near chemical bonding sites are important in the behavior of materials[1-3]. They influence crystal growth[4], decomposition[5], doping[6], adsorption[7], and catalytic reactivity[8]. Rodriguez et al. studies the characteristics and formation mechanisms of metallic bonds, providing guidance for the synthesis and surface modification of metallic materials[9]. Jiang et al. verified the relationship between the cohesive energy of nanocrystals and its size by using simulated and experimental data[10]. Bazant models covalent bonds in solids by inverting condensed energy curves and is very useful in understanding the fundamental properties of these materials[11]. The bond-order-length-strength (BOLS) correlation provides a good explanation of how atomic coordination contributes to the surface properties of a material[12,13]. Atomic systems located on surfaces have lower coordination numbers than those in bulk material and consequently different chemical and physical properties[14]. The atomic coordination number (CN) plays a key role of bond energy ($E_i$) and bond length ($d_i$), bonding-charge entrapment, and nonbonding polarization[15-17]. In density-functional theory (DFT), the electron density $\rho\left(\vec{r}\right)$ is the fundamental quantity for study[18-20]. Nevertheless, BOLS theory describes the relaxation of chemical bonds through the atomic CNs and does not establish a functional relationship between the electron density $\rho\left(\vec{r}\right)$ and bond energy $E_i$[21]. Therefore, it is necessary to establish a functional relationship between $\rho\left(\vec{r}\right)$ and $E_i$.

The binding-energy and bond-charge model (BBC) model is quantified chemical bonds by binding energy shift and deformation charge density. For the binding energy (BB) model, we use the central field approximation[22] and Tight-binding (TB) model[23], which can get the relationship between the energy shift and Hamiltonian. For the bond-charge (BC) model, we use the second-order term of energy expansion and use deformation charge density to calculate bonding states. For the dynamics of a potential function, we introduce the Laplace transform. We obtained the electronic structures and properties of 46 metal and semiconductor elements by DFT calculation. We combined BBC model[24] with the density-functional–tight-binding (DFTB) approach[25] and BOLS theory to establish a quantitative relationship between the electron density $\rho\left(\vec{r}\right)$ and bond energy $E_i$. The deformation charge density was calculated by DFT to obtain quantitative information about the interatomic deformation charge-bond energy $\Delta V_{\mathrm{bc}}\left(\vec{r}-\vec{r}^{'}\right)$, deformation atomic cohesive energy $\Delta E_{\mathrm{coh}}\left(i\right)$, and deformation bond energy density $\Delta E_{\mathrm{den}}\left(i\right)$. In the present study, we obtain better predictions by combining the BBC model with the BOLS theory and DFTB approaches.

## 2. Principles

### 2.1 DFT calculations

In this study, we applied a projection-based plane-wave DFT method[26-28], implemented through the Cambridge Sequential Total Energy Package (CASTEP) for material simulation at the atomic scale[29,30]. We applied General Gradient Approximate (GGA)-based Perdew-Burke-Ernzerhof (PBE) functionals for the calculations[31]. **Table S1** lists the lattice parameters of 46 metal and semiconductor elements; **Table S2** lists their cut-off energies and *k*-points. During the simulation, the energy converged to $10^{-6}$ eV, and the force applied to each atom converged to <0.01 eV/ $\dot{\mathrm{A}}$ [32].

### 2.2 DFTB and BBC model

The ground-state density is represented by a reference density, $\rho_0$, around which the actual density fluctuates: $\rho\left(\vec{r}\right)=\rho_0\left(\vec{r}\right)+\delta\rho\left(\vec{r}\right)$. The total energy $E\left[\rho_0+\delta\rho\right]$ is then written as a Taylor series to the second order:

$$E\left[\rho_0+\delta\rho\right]=E^0\left[\rho_0\right]+E^1\left[\rho_0+\delta\rho\right]+E^2\left[\rho_0,\left(\delta\rho\right)^2\right], \tag{1}$$

where

$$E^0[\rho_0]=\sum_i^N \varepsilon_i-\frac{1}{2}\iint \frac{\rho_0(\vec{r})\rho_0(\vec{r}')}{|\vec{r}-\vec{r}'|}d\vec{r}d\vec{r}'-\int V^{XC}[\rho_0]\rho_0(\vec{r})d\vec{r}+E^{XC}[\rho_0]$$

$$=\sum_i\langle\phi_i|-\frac{1}{2}\nabla^2+V_{eff}(\vec{r})|\phi_i\rangle-\frac{1}{2}\iint \frac{\rho_0(\vec{r})\rho_0(\vec{r}')}{|\vec{r}-\vec{r}'|}d\vec{r}d\vec{r}'-\int V^{XC}[\rho_0]\rho_0(\vec{r})d\vec{r}+E^{XC}[\rho_0] ,$$

(2)

$$E^1[\rho_0+\delta\rho]=Tr\left(\rho\hat{H}[\rho_0]\right)=\sum_i f_i\varepsilon_i ,$$

(3)

and

$$E^2\left[\rho_0,(\delta\rho)^2\right]=\frac{1}{2}\iint\left\{\frac{1}{|\vec{r}-\vec{r}'|}+\left.\frac{\delta^2E^{XC}}{\delta\rho(\vec{r})\delta\rho(\vec{r}')}\right|_{\rho_0}\right\}\delta\rho(\vec{r})\delta\rho(\vec{r}')d\vec{r}d\vec{r}'\approx\frac{1}{2}\iint\frac{1}{|\vec{r}-\vec{r}'|}\delta\rho(\vec{r})\delta\rho(\vec{r}')d\vec{r}d\vec{r}'$$

(4)

In **Eq. 2**, the $E^0[\rho_0]$ is called the "repulsive energy;" it determines the dispersion of the energy band. $V^{XC}$ and $E^{XC}$ are the potential and exchange correlation energies, respectively; they are typically fitted from *ab initio* calculations. In **Eq. 3**, $\varepsilon_i$ indicates the *i*th electronic level and $f_i$ the corresponding electronic occupation number. In case of non-self-consistent charge DFTB, $Tr\left(\rho\hat{H}[\rho_0]\right)=\sum_i f_i\varepsilon_i$, which makes the approach equivalent to a transferable tight-binding approximation. In **Eq. 4**, we ignore the effect of the second-order fluctuations $\delta^2E^{XC}$ of the exchange correlation energy. The $\rho_0(\vec{r})$ is the initial density function, and $\delta\rho(\vec{r})$ is the deformation density function.

As shown in **Fig. S1**, if we consider the effect of charge transfer on binding energy shifts, combining the BBC model with the DFTB model, we obtain:

$$E[\rho_0+\delta\rho]=E^0[\rho_0]+E^1[\rho_0+\delta\rho]+E^2\left[\rho_0,(\delta\rho)^2\right]$$
$$\approx\left(\sum_i\langle\phi_i|-\frac{1}{2}\nabla^2+V_{eff}(\vec{r})|\phi_i\rangle-\frac{1}{2}\iint\frac{\rho_0(\vec{r})\rho_0(\vec{r}')}{|\vec{r}-\vec{r}'|}\mathrm{d}\vec{r}\mathrm{d}\vec{r}'-\int V^{XC}[\rho_0]\rho_0(\vec{r})\mathrm{d}\vec{r}+E^{XC}[\rho_0]\right)+\sum_i f_i\varepsilon_i+\frac{1}{2}\iint\frac{1}{|\vec{r}-\vec{r}'|}\delta\rho(\vec{r})\delta\rho(\vec{r}')\mathrm{d}\vec{r}\mathrm{d}\vec{r}'$$

(5)

The second term $E^1[\rho_0+\delta\rho]$, which represents the contribution of the external field to the Hamiltonian, determines the relative energy-level shift $\Delta E'_v(x)$.Without the effect of the external field, the second term does not contribute to the Hamiltonian.

From the **Eq. 4,** the combined DFTB [33,34], we have:

$$\Delta V_{bc}(\vec{r}-\vec{r}')=\frac{1}{8\pi\varepsilon_0}\int d^3r\int d^3r'\frac{\delta\rho(\vec{r})\delta\rho(\vec{r}')}{|\vec{r}-\vec{r}'|},$$

(6)

where $\varepsilon_0$ is the dielectric constant of vacuum. Also, $\delta\rho$ satisfies the following relationship

$$(\delta\rho_{Hole\text{-}electron}\le\delta\rho_{Antibonding\text{-}electron}<\delta\rho_{No\ charge\ tranfer}=0<\delta\rho_{Nonbonding\text{-}electron}\le\delta\rho_{Bonding\text{-}electron})$$

(7)

For atomic (strong) bonding states:

$$\delta\rho_{Hole\text{-}electron}(\vec{r})\delta\rho_{Bonding\text{-}electron}(\vec{r}')<0(Strong\ Bonding).$$

(8)

For atomic nonbonding or weak-bonding states:

$$\begin{cases}\delta\rho_{Hole\text{-}electron}(\vec{r})\delta\rho_{Nonbonding\text{-}electron}(\vec{r}')<0(Nonbonding\ or\ Weak\ Bonding)\\ \delta\rho_{Antibonding\text{-}electron}(\vec{r})\delta\rho_{Bonding\text{-}electron}(\vec{r}')<0(Nonbonding\ or\ Weak\ Bonding)\\ \delta\rho_{Antibonding\text{-}electron}(\vec{r})\delta\rho_{Nonbonding\text{-}electron}(\vec{r}')<0(Nonbonding)\end{cases}$$

(9)

For atomic antibonding states:

$$\begin{cases} \delta\rho_{Nonbonding\text{-}electron}\left(\vec{r}\right)\delta\rho_{Bonding\text{-}electron}\left(\vec{r}'\right)>0(Antibonding) \\ \delta\rho_{Hole\text{-}electron}\left(\vec{r}\right)\delta\rho_{Antibonding\text{-}electron}\left(\vec{r}'\right)>0(Antibonding) \\ \delta\rho_{Hole\text{-}electron}\left(\vec{r}\right)\delta\rho_{Hole\text{-}electron}\left(\vec{r}'\right)>0(Antibonding) \\ \delta\rho_{Antibonding\text{-}electron}\left(\vec{r}\right)\delta\rho_{Antibonding\text{-}electron}\left(\vec{r}'\right)>0(Antibonding) \\ \delta\rho_{Nonbonding\text{-}electron}\left(\vec{r}\right)\delta\rho_{Nonbonding\text{-}electron}\left(\vec{r}'\right)>0(Antibonding) \\ \delta\rho_{Bonding\text{-}electron}\left(\vec{r}\right)\delta\rho_{Bonding\text{-}electron}\left(\vec{r}'\right)>0(Antibonding) \end{cases}$$

(10)

The formation of chemical bonds is related to fluctuations in electron density $\delta\rho$.

From the Hubbard model[35], we get

$$\begin{aligned} \hat{V}_{ee} &= \frac{1}{2}\int d^3r\int d^3r'\alpha_{\zeta}{}^{+}\left(\vec{r}\right)\alpha_{\zeta}\left(\vec{r}\right)V_{ee}(\vec{r}-\vec{r}')\alpha_{\zeta'}{}^{+}\left(\vec{r}'\right)\alpha_{\zeta'}\left(\vec{r}'\right) \\ &= \frac{1}{2\left|\vec{r}-\vec{r}'\right|}\int d^3r\int d^3r'\rho\left(\vec{r}\right)\rho\left(\vec{r}'\right) \end{aligned}$$

(11)

where $\alpha_{\zeta}\left(\vec{r}\right)$ and $\alpha_{\zeta'}{}^{+}\left(\vec{r}\right)$ are annihilation and creation operators, respectively. For the sake of completeness, we have endowed the electrons with a spin index, $\zeta = \uparrow/\downarrow$. The charge density is $\rho(\vec{r})=\alpha_{\zeta}^{+}\left(\vec{r}\right)\alpha_{\zeta}^{-}\left(\vec{r}\right)$. The quantity $V_{ee}(\vec{r}-\vec{r}')=\frac{1}{2\left|\vec{r}_i-\vec{r}_j'\right|}$ is the potential of the electron, which induces a transformation

$$\alpha_{\zeta}^{+}\left(\vec{r}\right)=\sum_{\vec{R}}\psi_{\vec{R}}^{*}\left(\vec{r}\right)a_{\vec{R}\zeta}^{+}\equiv\sum_{i}\psi_{\vec{R}i}^{*}\left(\vec{r}\right)a_{i\zeta}^{+} \quad .$$

(12)

Inserting (12) into (11) leads to the expansion $\hat{V}_{ee}=\sum_{ii'jj'}U_{ii'jj'}a_{i\zeta}^{+}a_{i'\zeta'}^{+}a_{j\zeta}^{-}a_{j'\zeta'}^{-}$, where

$$U_{ii'jj'}=\frac{1}{2}\int d^3r\int d^3r'\psi_{\vec{R}i}^{*}\left(\vec{r}\right)\psi_{\vec{R}j}\left(\vec{r}\right)V_{ee}(\vec{r}-\vec{r}')\psi_{\vec{R}i'}^{*}\left(\vec{r}'\right)\psi_{\vec{R}j'}\left(\vec{r}'\right)$$

(13)

is the Coulomb interaction.

Therefore, the electron density fluctuations can also be represented by the Hubbard model.

$$\Delta V_{bc}(\vec{r}-\vec{r}')=\frac{1}{4\pi\varepsilon_0}\times\frac{1}{2\left|\vec{r}-\vec{r}'\right|}\int d^3r\int d^3r'\delta\rho\left(\vec{r}\right)\delta\rho\left(\vec{r}'\right)$$

(14)

The deformation charge- bond energy $\Delta V_{bc}(\vec{r}-\vec{r}')$ is different from the Coulomb repulsion energy $\hat{V}_{ee}$. The deformation charge-bond energy considers the interaction between electrons and holes, and there are three cases of bonding, antibonding and nonbonding. The shielding factor $\mu$ is added to the equation.

**2.3 BBC model**

The TB model leads to the formulas

$$\begin{cases} H=-\dfrac{\hbar^2\nabla^2}{2m}+V_{atom}(\vec{r})+V_{cry}(\vec{r})(1+\Delta_{\mathrm{H}}) \\ V'_{cry}(\vec{r})=V_{cry}(\vec{r})(1+\Delta_{\mathrm{H}})=\gamma V_{cry}(\vec{r}) \end{cases},$$

(15)

$$\begin{cases} E_v(0)=-\left\langle v,i\right|-\dfrac{\hbar^2\nabla^2}{2m}+V_{atom}(\vec{r})\left|v,i\right\rangle \\ E_v(x)-E_v(0)=-\left\langle v,i\left|V'_{cry}(\vec{r})\right|v,i\right\rangle\left[1+\dfrac{\sum_j f(k)\left\langle v,i\left|V'_{cry}(\vec{r})\right|v,j\right\rangle}{\left\langle v,i\left|V'_{cry}(\vec{r})\right|v,j\right\rangle}\right] \\ \qquad =\gamma\alpha_v(1+\sum_j f(k)\dfrac{\beta_v}{\alpha_v})\approx\gamma\alpha_v\propto\left\langle E_i\right\rangle \end{cases},$$

(16)

$$\alpha_v=-\left\langle v,i\left|V_{cry}(\vec{r})\right|v,i\right\rangle,\beta_v=-\left\langle v,i\left|V_{cry}(\vec{r})\right|v,j\right\rangle$$

(17)

Here, $E_v(x)$ is the $v$-th energy level of a crystal atom, and $E_v(0)$ is the $v$-th energy level of an isolated atom; $V_{atom}(\vec{r})$ and $V_{cry}(\vec{r})$ are the potential energies of the atom and crystal. $\Delta_H$ is represents perturbation of crystal potential energy. $\gamma$ is represents bond energy ratio. Both exchange integral $\alpha_v$ and overlap integral $\beta_v$ in **Eq. 16** contribute to the energy band width. However, the energy band of the core of energy level is determined by $\alpha_v$, because $\beta_v$ is very small in the local energy band of the core of energy level, the term $\frac{\sum_j f(k)\cdot\beta_v}{\alpha_v} \ll 1$. $E_i$ represents the single bond energy. $|v,i\rangle$ represents the eigenwave function of an atom at the specific $i$th atomic site. The form of the periodic factor $f(k)$ is $e^{ik\vec{R}_j}$, where $k$ represents the wave vector. The $\vec{r}$ is the radius of the electron. $\vec{r}$ represents the electron coordinates and $\vec{R}$ represents the nuclei coordinates.

The inner electrons are affected by the potential energy of the crystal, which causes shifts in the core energy levels. Combining initial- and final-state effects with the TB model, we obtain[36]:

$$\begin{cases} V_{cry}(\vec{r})(1 + \Delta_H) = \gamma V_{cry}(\vec{r}) \\ V_{atom}(r) = -\frac{1}{4\pi\varepsilon_0}\frac{Z'e^2}{\vec{r}_i}, V_{cry}(\vec{r}) = -\sum_{i,j,\vec{R}_j\neq 0}\frac{1}{4\pi\varepsilon_0}\frac{Z'e^2}{\left|\vec{r}_i - \vec{R}_j\right|} \end{cases}, \tag{18}$$

The core electron will not only be attracted by the nuclear charge, but also be excluded by other electrons. The repulsive effect of electrons will reduce the attractive effect of the nucleus. The effective positive charge of the ion is $Z' = Z - \sigma$, considering the charge shielding effect $\sigma$, where $Z$ is the nuclear charge. $\sigma = \left|\vec{r}_i - \vec{R}_j\right| / \left|\vec{r}_i - \vec{r}_j\right|$ can be written as a Hamiltonian containing electron interaction terms. However, considering the repulsion of electrons, the effective positive charge number $Z'$ of the ion can be varied, as shown in **Fig. 1**.

$$E_v(x) - E_v(0) \cong -\langle v,i|V_{cry}(\vec{r})(1 + \Delta_H)|v,i\rangle = -(1+\delta\gamma)\langle v,i|V_{cry}(\vec{r})|v,i\rangle \tag{19}$$

$$E_v(x) - E_v(B) \cong -\delta\gamma\langle v,i|V_{cry}(\vec{r})|v,i\rangle$$

(20)

$\delta\gamma=\gamma-1$ is relative bond energy ratio and $B$ indicates bulk atoms. $\Delta E_v(B)$ is represents the energy shift of an atom in an ideal bulk.

The energy-level shifts in an external field obtained from Eqs. **18**, **19** and **20** are:

$$\delta\gamma=-1(Antibonding),\ \Delta E_v\left(x\right)=\Delta E_v\left(0\right)=0 \qquad \text{(isolated atom,neutral atom)}$$

(21)

$$\delta\gamma=0(Bonding),\Delta E_v\left(x\right)=\Delta E_v\left(B\right)=-(-\sum_{l,\vec{R}_l\neq 0}\langle v,i|\frac{1}{4\pi\varepsilon_0}\frac{Z'e^2}{\left|\vec{r}-\vec{R}_l\right|}|v,i\rangle)>0 \quad \text{(bulk atoms,core level loses electrons)}$$

(22)

$$\delta\gamma>0\begin{cases}\delta\gamma>1(Bonding),\Delta E_v\left(x\right)=-(-\langle v,i|\sum\limits_{l,\vec{R}_l\neq 0}\frac{(1+\delta\gamma)}{4\pi\varepsilon_0}\frac{Z'e^2}{\left|\vec{r}-\vec{R}_l\right|}|v,i\rangle)>0 \quad \text{(potential well becomes deeper)}\\ 1>\delta\gamma>0(Bonding),\ \Delta E_v\left(x\right)=-(-\langle v,i|\sum\limits_{l,\vec{R}_l\neq 0}\frac{(1+\delta\gamma)}{4\pi\varepsilon_0}\frac{Z'e^2}{\left|\vec{r}-\vec{R}_l\right|}|v,i\rangle)>0 \quad \text{(core level loses electrons)}\end{cases}$$

(23)

$$\delta\gamma<0\begin{cases}-1<\delta\gamma<0(Nonbonding),\ \Delta E_v\left(x\right)=-(-\sum\limits_{l,\vec{R}_l\neq 0}\langle v,i|\frac{(1+\delta\gamma)}{4\pi\varepsilon_0}\frac{Z'e^2}{\left|\vec{r}-\vec{R}_l\right|}|v,i\rangle)>0 \text{ (core level gets electrons)}\\ \delta\gamma<-1(Antibonding),\Delta E_v\left(x\right)=-(-\sum\limits_{l,\vec{R}_l\neq 0}\langle v,i|\frac{(1+\delta\gamma)}{4\pi\varepsilon_0}\frac{Z'e^2}{\left|\vec{r}-\vec{R}_l\right|}|v,i\rangle<0) \text{ ( potential well becomes shallower or barrier potential)}\end{cases}$$

(24)

The core of energy level shift determines the transfer of electrons between core energy levels. For under-coordination system, the positive and negative values of $\delta\gamma$ represent the positive and negative energy shifts. For elemental elements, the core of energy level gains electron ($\delta\gamma<0$), the core band widens, the atomic coordination number increases (up to the bulk coordination number), and the bond energy weakens. The core level loses electron ($\delta\gamma>0$), the core band becomes narrower, the atomic coordination number decreases, and the bond energy is strengthened. For compound, hetero-coordination system. When $\delta\gamma<0$, it means that the core of energy level is negatively shifted, the impurity level appears in the core band, and the core level gains electrons. The potential energy ( $\gamma V_{cry}(\vec{r})$ ) after perturbation is weakened, and the chemical bond becomes weaker. On the contrary, when $\delta\gamma>0$, it indicates that the core of energy level is positively shifted, impurity levels appear in the core band, electrons are lost in the core of energy level, and the potential energy ( $\gamma V_{cry}(\vec{r})$ ) after perturbation is enhanced, and the chemical bond is enhanced. When $\delta\gamma > 1$, the contribution $\delta\gamma V_{cry}(\vec{r})$ of the external field to the potential energy is greater than $V_{cry}(\vec{r})$ ($\delta\gamma V_{cry}(\vec{r}) > V_{cry}(\vec{r})$) of the crystal potential energy itself, the core band forms the impurity level and forms A new potential well (compound). $\delta\gamma < -1$, the potential energy $\gamma V_{cry}(\vec{r})$ after perturbation is less than 0, and the core band forms the impurity level, forming new potential well (compound) or barrier.

In the BBC model, the occupied states of the electron orbital are related to the chemical bonds of the atoms. There are three electron states of atomic bonding of core band: antibonding-electron states, nonbonding-electron states and bonding-electron states. **Fig. 2** shows a schematic of BBC model.

Considering the dynamic process of potential function:

$$\begin{cases} V_i(\vec{r},t) = (1+f(t))V_0(\vec{r}) \\ \mathcal{L}(f(t)) = F(s) = \int_{-\infty}^{+\infty} f(t)e^{-st}dt \\ f(t) = \mathcal{L}^{-1}(F(s)), \hat{S}_t = e^{-iEt/\hbar} = e^{-iwt} \end{cases} \tag{25}$$

$\mathcal{L}$ is Laplace transform. $\hat{S}_t$ is an operator. The $s=\kappa+jw$ is a complex number. When $\kappa$ is 0, the function $f(t)$ is Fourier transform. **Table 1** is the Laplace transform formula.

Energy-level shifts reflect bond changes and electron transfers. Through DFT calculations and X-ray photoelectron spectroscopy (XPS) analysis, we can pass the known reference values $\Delta E_v'(x) = E_v(x) - E_v(B)$, $\Delta E_v(B) = E_v(B) - E_v(0)$, and $\Delta E_v(x) = E_v(x) - E_v(0)$ to calculate the binding-energy ratio $\gamma$ induced by the external field:

$$\gamma = \frac{E_v(x) - E_v(0)}{E_v(B) - E_v(0)} \approx \frac{Z_x}{Z_b}\frac{d_b}{d_x} = \left(\frac{Z_b - \sigma'}{Z_b}\right)\frac{d_b}{d_x} = \left(\frac{d_x}{d_b}\right)^{-m},$$

(26)

$\sigma'$ is the relative charge shielding factor. *m* is an indicator for the bond nature of a specific material. The *m* is related to the charge shielding factor $\sigma'$, $m=1-\frac{In\frac{Z_b-\sigma'}{Z_b}}{In\left(\frac{d_x}{d_b}\right)}$.

For occupied electrons at the same energy level, $\sigma'$ is a very small value, so the charge screening effect can be ignored, *m* = 1. The properties of chemical bonds can be obtained in terms of the binding-energy ratio $\gamma$. Combining the BOLS correlation, we obtain:

$$\begin{cases}\delta E_{den}(i)=(E_i/d_i^3)/(E_b/d_b^3)-1=\gamma^{3+m}-1 & \text{(relative bond energy density)}\\ \delta E_{coh}(i)=z_iE_i/z_bE_b-1=z_{ib}\gamma^m-1 & \text{(relative atomic cohesive energy)}\end{cases}$$

(27)

$d_i$ and $E_i$ are the length and energy of the bond, respectively. $E_b$ represents the bond energy in the ideal bulk. $d_b$ represents the bond length in the ideal bulk. $E_{\mathrm{coh}}(i)$ is the cohesive energy of the atoms and $E_{den}(i)$ is the bond energy density.

## 3. Results and discussion

The geometric initial structures of the 46 metal and semiconductor elements are shown in **Fig. S2**, and their lattice constants in **Table S1**. It can be seen from **Fig. S2** that C, Si, Ge, and Sn have diamond structures; Li, Na, K, V, Cr, Fe, Rb, Nb, Mo, In, Cs, Ba, Ta, and W have body-centered cubic (bcc) structures; Be, Mg, Sc, Ti, Zn, Y, Zr, Tc, Ru, Cd, Lu, Co, Hf, Re, Os, and Tl have hexagonal close-packed (hcp) structures; and Al, Ca, Ni, Cu, Sr, Rh, Pd, Ag, Ir, Pt, Au, and Pb have faced-centered cubic (fcc) structures. The bonds formed by these elements are metallic. The diamond-structure elements form covalent bonds through the valence electrons of surrounding atoms. Elements with the same lattice structure tend to have similar bonding states. The Mn, Ga and Hg are not the bcc, hcp, fcc and diamond structures, so they are not included in the DFT calculation in the paper.

We used DFT to calculate the partial density of states of 46 metal and semiconductor elements, as is shown in the **Fig. S3**. The band gaps of C, Si, and Ge is shown in **Fig. S4**. We used PBE, LDA and HSE06 functional calculations. In Table S3, HSE06 functional was used for calculation, the band-gap value was close to the experimental result[37,38]. The DFT was used to calculate $\delta\rho(r)$ for 46 metal and semiconductor elements (**Fig. 3**), as is shown in **Table 2**. The values of the differential charge density diagram are shown in **Fig.S5**. Here, we analyze the electron bonding states using the BBC model. From the charge transfer process of the deformation charge-density diagram, we obtain the electronic states of chemical bonds of various elements. Given $\delta\rho(r)$, we use the BBC model to obtain the relevant bonding value. **Table 2** lists the bonding-state deformation charge densities $\delta\rho^{Bonding-electron}(\vec{r}')\left(e/\mathring{A}^3\right)$ of each element determined by DFT calculation, the atomic bond length *d*, and $\Delta V_{\mathrm{bc}}(\vec{r}-\vec{r}')$, where $\Delta V_{\mathrm{bc}}(\vec{r}-\vec{r}')$ is the potential energy of electron exchange. **Eq. 6** describes the relationship between $\delta\rho(r)$ and $\Delta V_{\mathrm{bc}}(\vec{r}-\vec{r}')$; for crystal structures, the latter depends on the electron densities of the different bonding electronic states. In **Fig. 4,** we show how $\Delta V_{\mathrm{bc}}(\vec{r}-\vec{r}')$ changed for the 46 metals and semi-metals across the periodic table. The density of the electrons is related to the correlation exchange functional used by DFT calculation. The electron density obtained by the BBC model is related to the correlation exchange functional, and the values obtained by different correlation exchange functional will be different.

The deformation charge- bond energy can be written in the form $\vec{r}_{ij}$ as follows:

$$\Delta V_{bc}(\vec{r}_{ij}) = \frac{1}{8\pi\varepsilon_0}\int d^3r_i\int d^3r_j\frac{\delta\rho(\vec{r}_i)\delta\rho(\vec{r}_j)}{|\vec{r}_i-\vec{r}_j|} = \frac{1}{8\pi\varepsilon_0}\int d^3r_i\int d^3r_j\frac{\delta\rho(\vec{r}_i)\delta\rho(\vec{r}_j)}{|\vec{r}_{ij}|} \tag{28}$$

The bond length can be approximated as $d_{ij}\approx 2|\vec{r}_{ij}|$, and we get further:

$$\Delta V_{bc}(\vec{r}_{ij})\approx\Delta E_i = \int d^3r\int d^3r'\frac{\delta\rho(\vec{r})\delta\rho(\vec{r}')}{4\pi\varepsilon_0 d_{ij}}$$

(29)

From **Eq. 6, Eq. 28** and **Eq. 29**, using to the $\delta\rho(r)$ obtained from the DFT calculation, we can calculate:

$$\begin{cases} \Delta E_{coh}(i) = z_i \Delta E_i = z_i \int d^3r \int d^3r' \frac{\delta\rho(\vec{r})\delta\rho(\vec{r}')}{4\pi\varepsilon_0 d_i} = \frac{z_i}{4\pi\varepsilon_0 d_i} \int d^3r \int d^3r' \delta\rho(\vec{r})\delta\rho(\vec{r}') \\ \Delta E_{den}(i) = \frac{\Delta E_i}{d_i^3} = \frac{1}{d_i^3} \int d^3r \int d^3r' \frac{\delta\rho(\vec{r})\delta\rho(\vec{r}')}{4\pi\varepsilon_0 d_i} = \frac{1}{4\pi\varepsilon_0 d_i^4} \int d^3r \int d^3r' \delta\rho(\vec{r})\delta\rho(\vec{r}') \end{cases}$$

(30)

From **Eq. 30,** we obtained the results shown in **Table 3**. **Fig. 5** shows $\Delta E_{\mathrm{coh}}(i)$ and $\Delta E_{\mathrm{den}}(i)$ for the 46 metal and semiconductor elements arranged by atomic number. The sign of $\Delta E_{\mathrm{coh}}(i)$ for the atoms determines the thermal stability[39]. In addition, we compared the $\Delta E_{\mathrm{coh}}(i)$ and cohesive energies of metals and semi-metals as is show in **Fig. 5a.** The value of $\Delta E_{\mathrm{den}}(i)$ indicates that the mechanical properties of the material[40]. We compared the $\Delta E_{\mathrm{den}}(i)$ and atomic concentration of metals and semi-metals, as is show in **Fig. 5b.** Philipsen et al. calculated the cohesive energy of Sc, Ti, V, Cr, Fe, Co, Ni, Cu and Zn metal using formula $E_{coh}=E_{\mathrm{crystal}}-E_{atoms}$. In **Fig. 6** and **Table 4**, we calculated atomic cohesive energy using formula $\Delta E_{coh}(i) = z_i \Delta E_i$ , the results showing the same trend of change.[41-43]. For BC model, it comes from the second-order term of energy expansion, which can only be compared with trends in experimental data, and the trend is basically the same. We find that the calculated deformation atomic cohesive energy $\Delta E_{\mathrm{coh}}(i)$ and deformation bond energy density $\Delta E_{\mathrm{den}}(i)$ are consistent with the trend of cohesive energies and atomic concentration. Atomic concentration and cohesive energies value are from the ref[44]. Considering the complex electronic structure system of heavy elements, we did not calculate the $\Delta E_{\mathrm{den}}(i)$ and $\Delta E_{\mathrm{den}}(i)$ of the heavy element system. The BBC model agrees with the DFT calculations [45] of atomic cohesive energy $E_{\mathrm{coh}}(i)$. The smaller the lattice size (the smaller the atomic radius), the stronger the atomic cohesive energy.

## 4. Conclusions

The BBC model is quantified chemical bonds by binding energy shift and deformation charge density. The BB model in the BBC model was derived from the BOLS model expanding model. Considering the electronic shielding effect, the calculated data of DFT were basically consistent with the experimental data of XPS[46]. For BC model, it comes from the second-order term of energy expansion, which can only be compared with trends in experimental data, and the trend is basically the same. The BBC model provided a quantitative relationship between the electron density and of bond energy. Furthermore, we obtained quantitative information about deformation charge-bond energy $\Delta V_{\mathrm{bc}}\left(\vec{r}-\vec{r}'\right)$, which determines the electronic bonding state; $E_{\mathrm{coh}}\left(i\right)$, which determines the thermal stability of the material; and $E_{\mathrm{den}}\left(i\right)$, which determines the mechanical properties. This can provide better theoretical understanding of chemical bonding states in metal and semiconductor elements.

## Figure and Table Captions

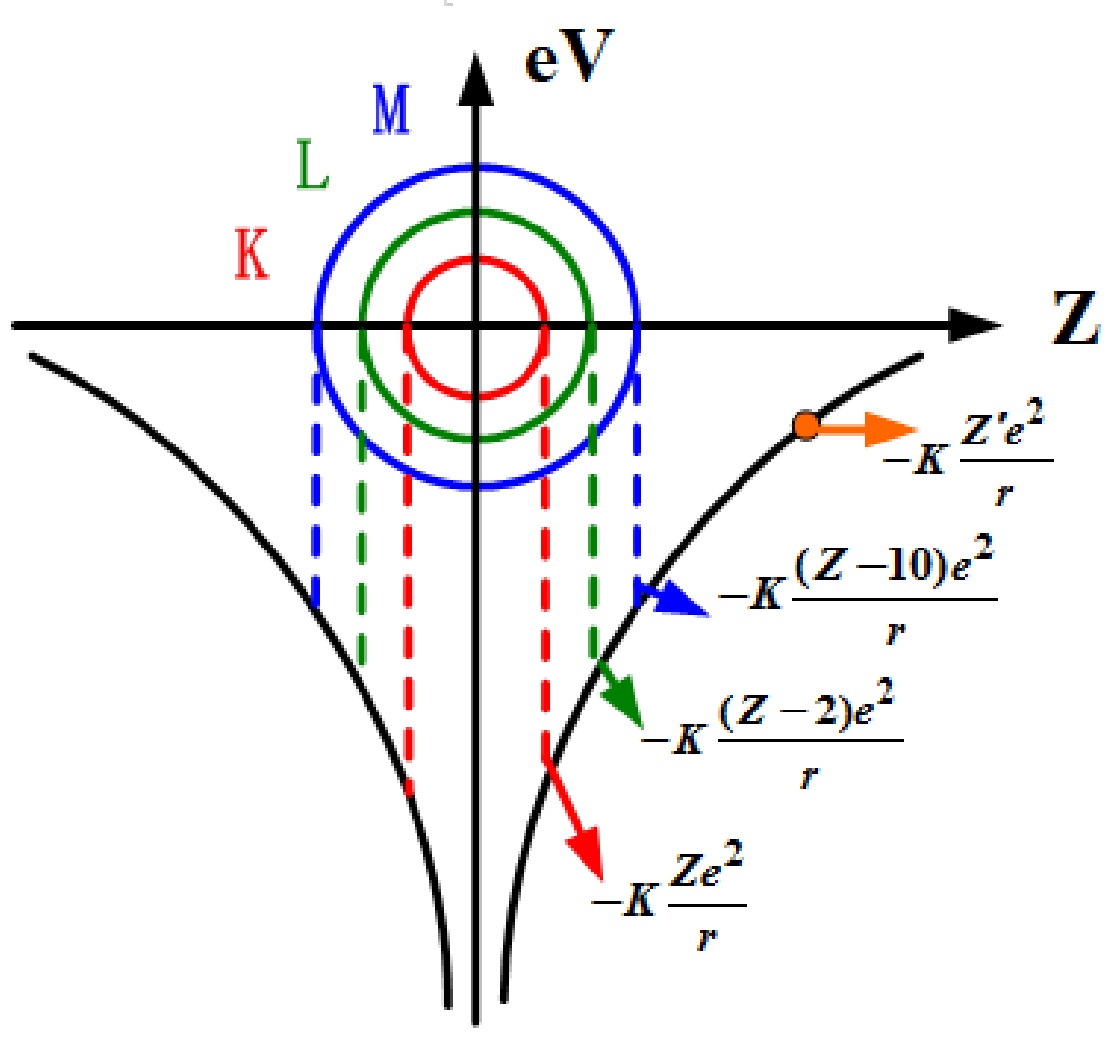

**Fig. 1** Schematic diagram of the atom potential energy, $K=\frac{1}{4\pi\varepsilon_0}$.

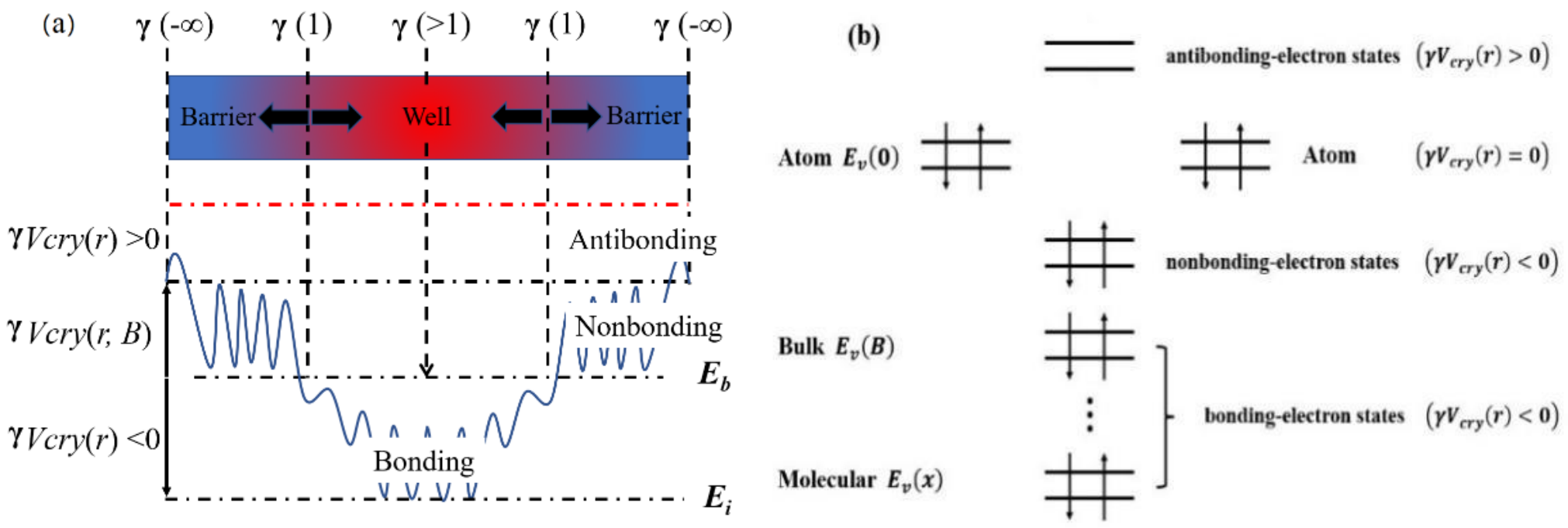

**Fig. 2** Schematic showing the BBC model.

(a)

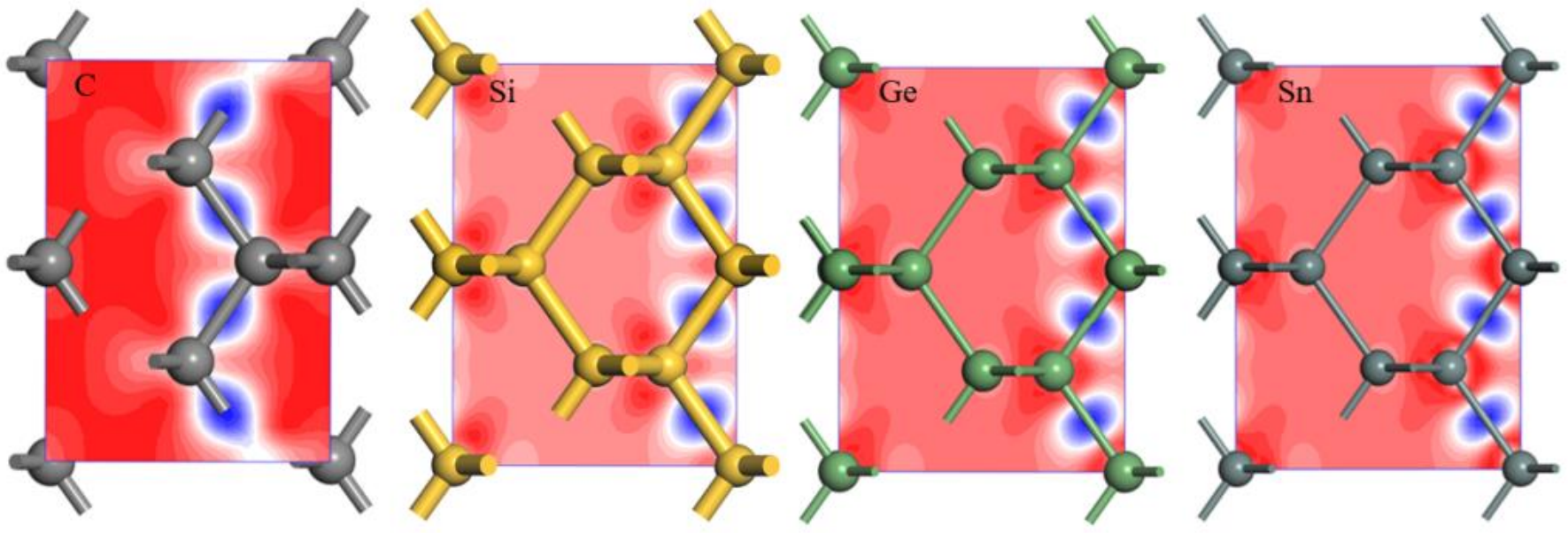

**(b)**

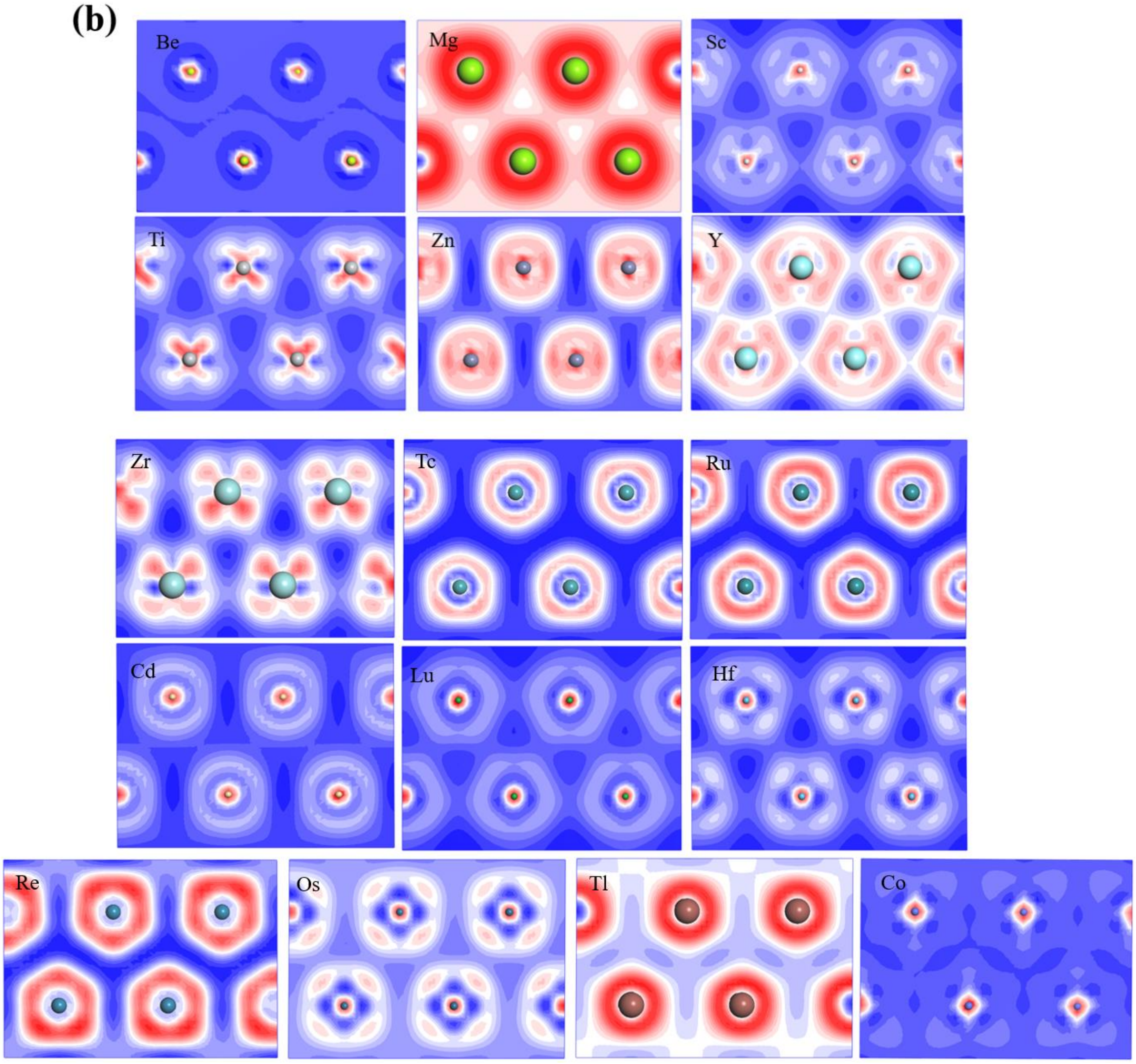

**(c)**

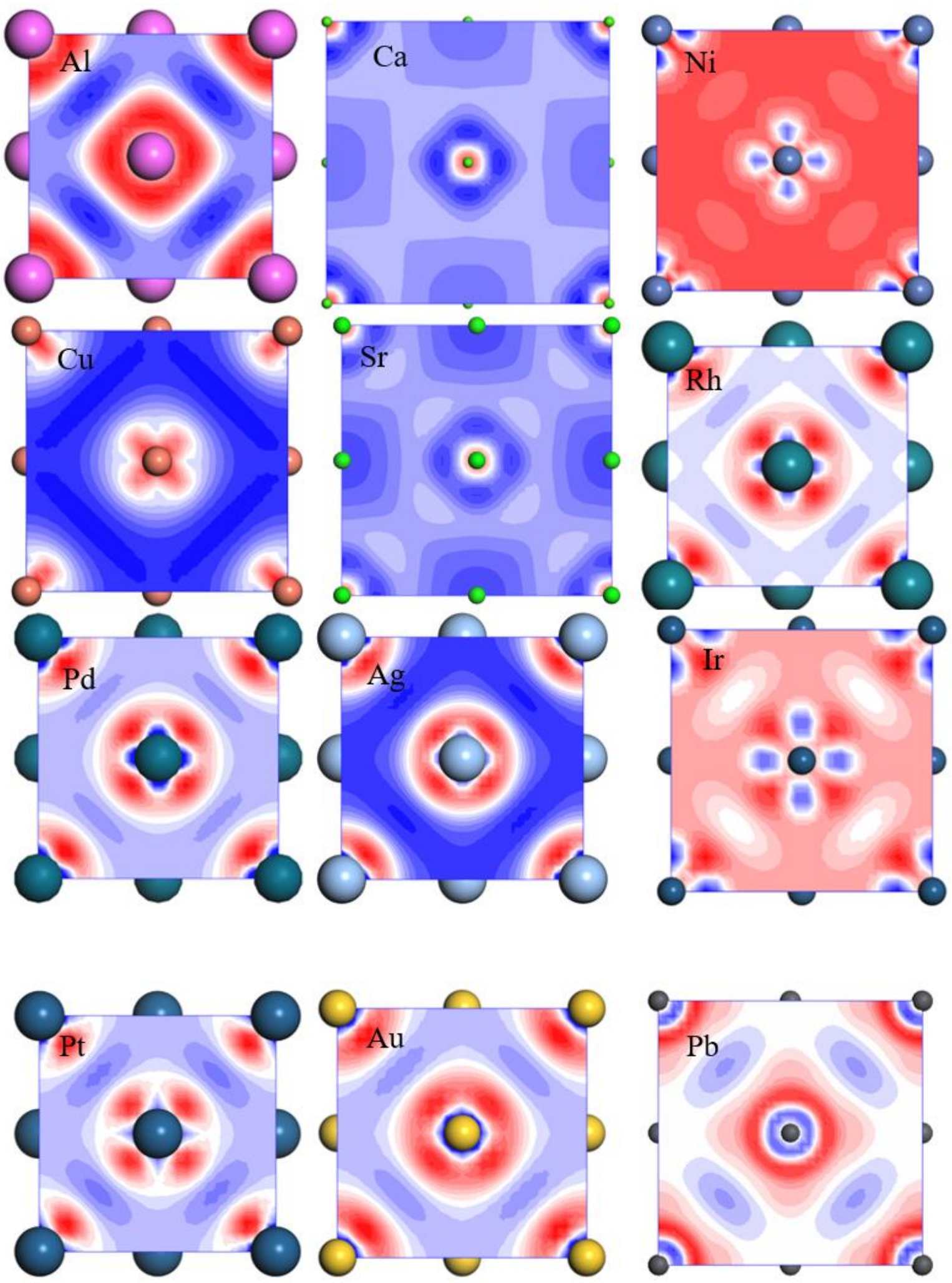

**(d)**

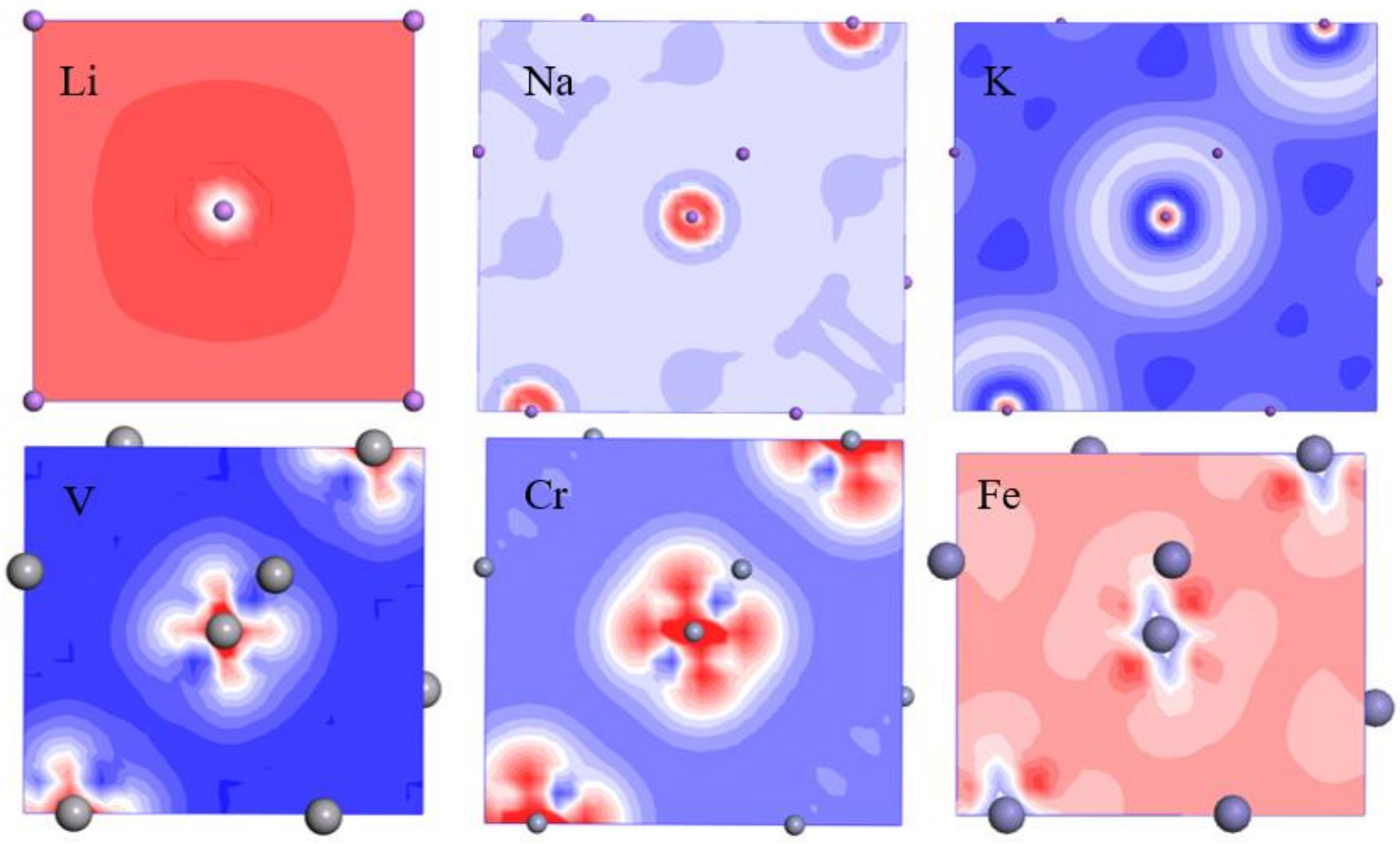

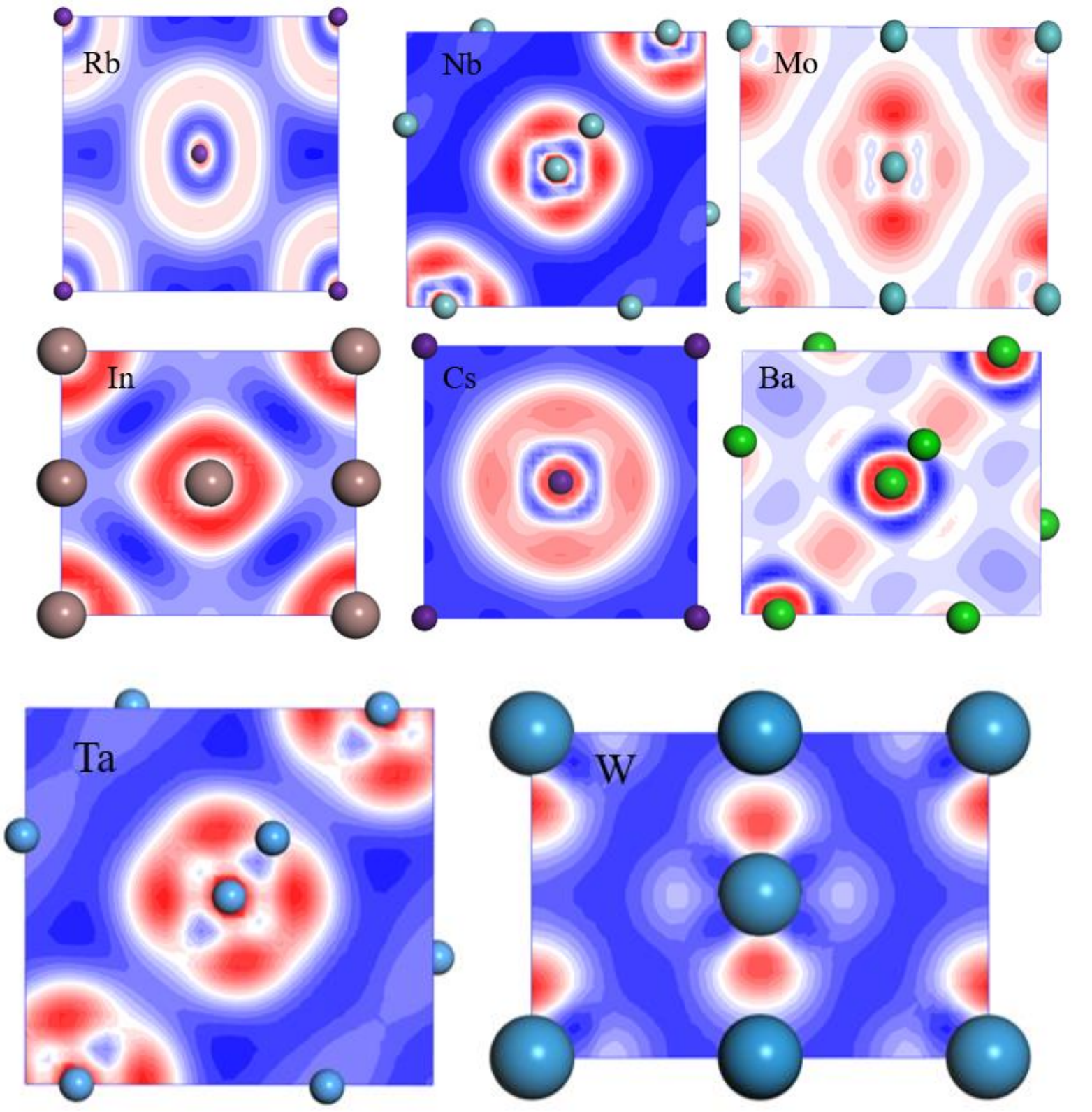

**Fig. 3** Deformation charge-density diagrams of (a) diamond-structure elements, (b) hcp structure elements, (c) fcc structure elements, and (d) bcc structure elements. The blue and red areas represent the concentration and decrease of electrons, respectively.

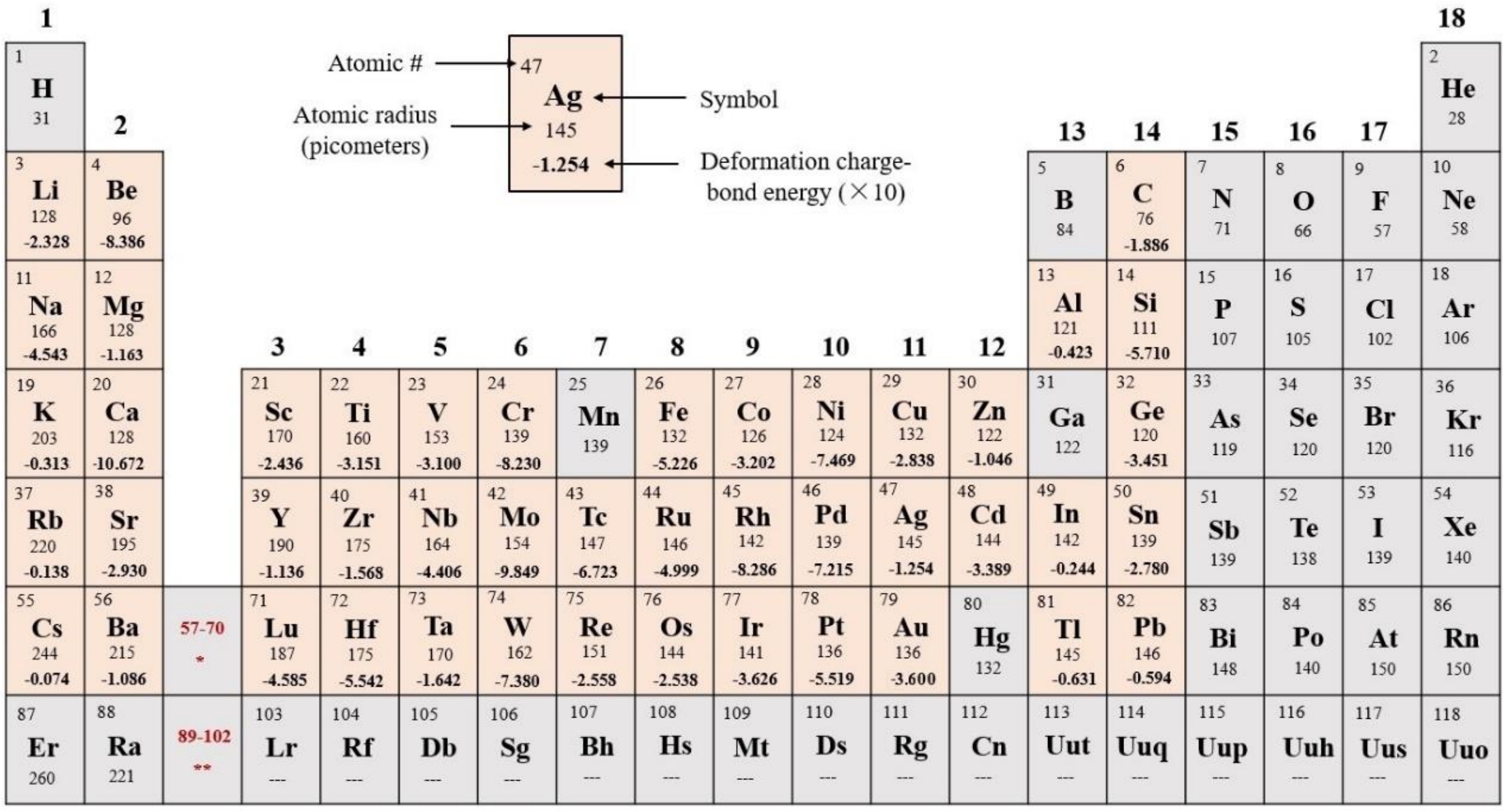

**Fig. 4** The deformation charge-bond energy $\Delta V_{bc}\left(\vec{r}-\vec{r}'\right)$ values of 46 metal and semiconductor elements and semi-metals, with their positions in the periodic table.

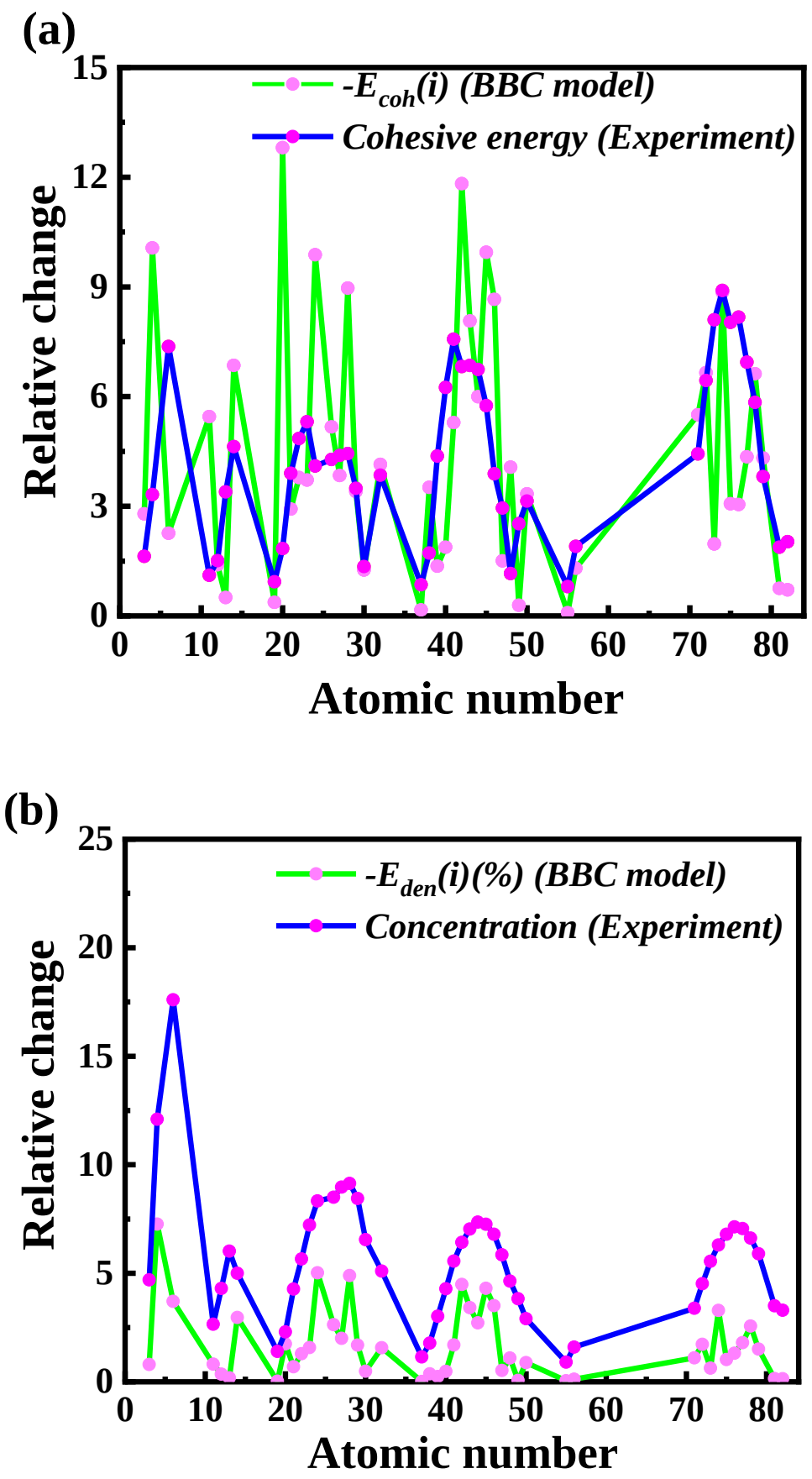

**Fig. 5** (a) deformation atomic cohesive energy $-\Delta E_{\text{coh}}(i)$ (eV/atom) and atomic cohesive energy(eV/atom), (b) deformation bond energy density $-\Delta E_{\text{den}}(i)$ (eV•$\dot{\text{A}}^{-3}$) and atomic concentration($10^{28}m^{-3}$). Atomic concentration and cohesive energies value are from the ref[44].

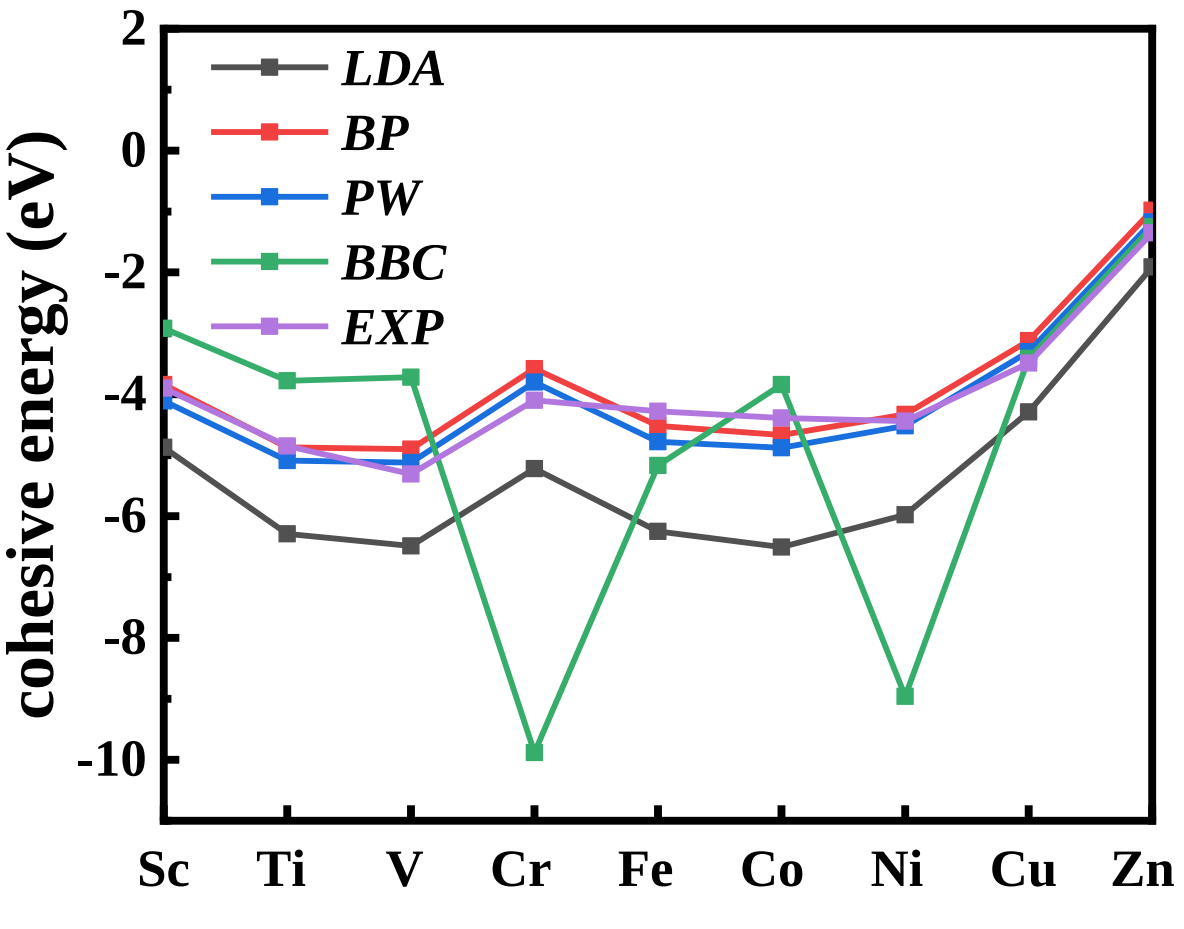

Fig. 6 The cohesive energy calculated by the BBC model are compared with the experimental and theoretical values.

**Table 1** Laplace transform formula.

| Primitive function $f(t)=\mathcal{L}^{-1}[F(s)]$ | Laplace transform $F(s)=\mathcal{L}[f(t)]$ | Convergence region |
|---|---|---|
| $\delta(t)$ | $1$ | $\infty > s > -\infty$ |
| $1$ | $\frac{1}{s}$ | $s>0$ |
| $e^{-at}$ | $\frac{1}{s+a}$ | $s>-a$ |
| $\sin(at)$ | $\frac{a}{s^2+a^2}$ | $s>0$ |
| $\cos(at)$ | $\frac{s}{s^2+a^2}$ | $s>0$ |
| $\sinh(at)$ | $\frac{a}{s^2-a^2}$ | $s>\|a\|$ |
| $\cosh(at)$ | $\frac{s}{s^2-a^2}$ | $s>\|a\|$ |
| $e^{at}\sin(bt)$ | $\frac{b}{(s-a^2)+b^2}$ | $s>a$ |
| $e^{at}\cos(bt)$ | $\frac{s-a}{(s-a^2)+b^2}$ | $s>a$ |
| $e^{at}\sinh(bt)$ | $\frac{b}{(s-a^2)-b^2}$ | $s-a>\|b\|$ |
| $e^{at}\cos(bt)$ | $\frac{s-a}{(s-a^2)-b^2}$ | $s-a>\|b\|$ |

**Table 2** BBC model values of deformation charge density $\delta\rho(\vec{r})$ and deformation charge-bond energy $\Delta V_{bc}(\vec{r}-\vec{r}')$

（$\varepsilon_0 = 8.85\times10^{-12}C^2N^{-1}\mathrm{m}^{-2}, e=1.60\times10^{-19}C, |\vec{r}-\vec{r}'| = d_i/2 \approx r_s$）

| Element | Bonding-electron density $\delta\rho(r)\left(e/\mathring{A}^3\right)$ | Hole-electron density $\delta\rho(r')\left(e/\mathring{A}^3\right)$ | Electronic radius $r_s(\mathring{A})$ | Deformation charge-bond energy $\Delta V_{bc}(\vec{r}-\vec{r}')(\mathrm{eV})$ |
|---|---|---|---|---|
| Li | 0.1460 | -0.0256 | 1.54 | -0.2328 |
| Na | 0.0235 | -0.1060 | 1.91 | -0.4543 |
| K | 0.0050 | -0.0124 | 2.34 | -0.0313 |
| Rb | 0.0027 | -0.0073 | 2.50 | -0.0138 |
| Cs | 0.0022 | -0.0032 | 2.71 | -0.0074 |
| Be | 0.1011 | -0.6261 | 1.13 | -0.8386 |
| Mg | 0.0542 | -0.0285 | 1.60 | -0.1163 |
| Ca | 0.0526 | -0.0951 | 1.97 | -1.0672 |
| Sr | 0.0192 | -0.0462 | 2.15 | -0.2930 |
| Ba | 0.0137 | -0.0195 | 2.24 | -0.1086 |
| Sc | 0.0250 | -0.1141 | 1.64 | -0.2436 |
| Y | 0.0226 | -0.0371 | 1.80 | -0.1136 |
| Lu | 0.0280 | -0.1471 | 1.73 | -0.4585 |
| Ti | 0.0415 | -0.1646 | 1.45 | -0.3151 |
| Zr | 0.0275 | -0.0758 | 1.60 | -0.1568 |
| Hf | 0.0398 | -0.1905 | 1.59 | -0.5542 |
| V | 0.0462 | -0.2080 | 1.35 | -0.3100 |
| Nb | 0.0596 | -0.1447 | 1.48 | -0.4406 |
| Ta | 0.0444 | -0.0724 | 1.48 | -0.1642 |
| Cr | 0.0955 | -0.3628 | 1.27 | -0.8230 |
| Mo | 0.0763 | -0.3338 | 1.40 | -0.9849 |
| W | 0.0733 | -0.2513 | 1.41 | -0.7380 |
| Tc | 0.0726 | -0.2872 | 1.35 | -0.6723 |
| Re | 0.0549 | -0.0977 | 1.46 | -0.2558 |
| Fe | 0.1030 | -0.1761 | 1.27 | -0.5226 |
| Ru | 0.0763 | -0.2275 | 1.32 | -0.4999 |
| Os | 0.0634 | -0.1289 | 1.34 | -0.2538 |
| Co | 0.0519 | -0.2701 | 1.26 | -0.3202 |
| Rh | 0.1618 | -0.1648 | 1.34 | -0.8286 |
| Ir | 0.1401 | -0.0774 | 1.36 | -0.3626 |
| Ni | 0.3864 | -0.0917 | 1.24 | -0.7469 |
| Pd | 0.1011 | -0.2056 | 1.37 | -0.7215 |
| Pt | 0.1040 | -0.1423 | 1.39 | -0.5519 |
| Cu | 0.0458 | -0.2508 | 1.28 | -0.2838 |
| Ag | 0.0301 | -0.0935 | 1.44 | -0.1254 |
| Au | 0.0814 | -0.0993 | 1.44 | -0.3600 |
| Zn | 0.0323 | -0.0869 | 1.39 | -0.1046 |
| Cd | 0.0304 | -0.1627 | 1.57 | -0.3389 |
| Al | 0.0209 | -0.0470 | 1.43 | -0.0423 |
| In | 0.0102 | -0.0264 | 1.66 | -0.0244 |
| Tl | 0.0232 | -0.0244 | 1.73 | -0.0631 |
| C | 0.4330 | -0.1287 | 0.86 | -0.1886 |
| Si | 0.1857 | -0.0990 | 1.34 | -0.5710 |
| Ge | 0.1487 | -0.0600 | 1.40 | -0.3451 |
| Sn | 0.1001 | -0.0392 | 1.58 | -0.2780 |
| Pb | 0.0184 | -0.0273 | 1.75 | -0.0594 |

**Table 3** Deformation atomic cohesive energies $\Delta E_{\text{coh}}(i)$ and deformation bonding energy densities $\Delta E_{\text{den}}(i)$ from **Eq. 30**.

| **Element** | $\Delta E_{\text{coh}}(i)(\text{eV/atom})$ | $\Delta E_{\text{den}}(i)\left(\text{eV}\bullet\dot{\text{A}}^{-3}\right)$ |
|---|---|---|
| Li | -2.7932 | -0.00797 |
| Na | -5.4516 | -0.00815 |
| K | -0.3753 | -0.00031 |
| Rb | -0.1661 | -0.00011 |
| Cs | -0.0888 | -0.00046 |
| Be | -10.0637 | -0.07265 |
| Mg | -1.3955 | -0.00355 |
| Ca | -12.8059 | -0.01745 |
| Sr | -3.5160 | -0.00369 |
| Ba | -1.3030 | -0.00121 |
| Sc | -2.9232 | -0.00690 |
| Y | -1.3630 | -0.00243 |
| Lu | -5.5025 | -0.01107 |
| Ti | -3.7818 | -0.01292 |
| Zr | -1.8822 | -0.00479 |
| Hf | -6.6501 | -0.01723 |
| V | -3.7196 | -0.01575 |
| Nb | -5.2867 | -0.01699 |
| Ta | -1.9710 | -0.00633 |
| Cr | -9.8765 | -0.05022 |
| Mo | -11.8189 | -0.04487 |
| W | -8.8554 | -0.03291 |
| Tc | -8.0677 | -0.03416 |
| Re | -3.0701 | -0.01028 |
| Fe | -5.1704 | -0.02629 |
| Ru | -5.9987 | -0.02717 |
| Os | -3.0456 | -0.01319 |
| Co | -3.8423 | -0.02001 |
| Rh | -9.9433 | -0.04305 |
| Ir | -4.3513 | -0.01802 |
| Ni | -8.9626 | -0.04897 |
| Pd | -8.6580 | -0.03507 |
| Pt | -6.6223 | -0.02569 |
| Cu | -3.4061 | -0.01692 |
| Ag | -1.5051 | -0.00525 |
| Au | -4.3182 | -0.01506 |
| Zn | -1.2547 | -0.00487 |
| Cd | -4.0666 | -0.01095 |
| Al | -0.5072 | -0.00181 |
| In | -0.2929 | -0.00067 |
| Tl | -0.7568 | -0.00152 |
| C | -2.2628 | -0.03706 |
| Si | -6.8520 | -0.02966 |
| Ge | -4.1414 | -0.01572 |
| Sn | -3.3359 | -0.00881 |
| Pb | -0.7126 | -0.00139 |

**Table 4** For Sc, Ti, V, Cr, Fe, Co, Ni, Cu and Zn elements, the calculated results (eV) are compared with the experimental values (eV) using different functional methods.

| EI | LDA | BP | PW | BBC | EXP |
|---|---|---|---|---|---|
| Sc | -4.87 | -3.84 | -4.11 | -2.92 | -3.90 |
| Ti | -6.29 | -4.87 | -5.09 | -3.78 | -4.85 |
| V | -6.49 | -4.90 | -5.12 | -3.72 | -5.31 |
| Cr | -5.22 | -3.58 | -3.8 | -9.88 | -4.1 |
| Fe | -6.25 | -4.52 | -4.78 | -5.17 | -4.28 |
| Co | -6.51 | -4.67 | -4.88 | -3.84 | -4.39 |
| Ni | -5.98 | -4.33 | -4.52 | -8.96 | -4.44 |
| Cu | -4.29 | -3.12 | -3.30 | -3.41 | -3.49 |
| Zn | -1.91 | -0.98 | -1.17 | -1.25 | -1.35 |

# Supplemental Material

# Atomic bonding states of metal and semiconductor elements

Liangjing Ge, Maolin Bo*

Key Laboratory of Extraordinary Bond Engineering and Advanced Materials Technology (EBEAM) of Chongqing, Yangtze Normal University, Chongqing 408100,

China

*Corresponding Author: E-mail: bmlwd@yznu.edu.cn (Maolin Bo)

## 2.1 Supporting materials of BBC model

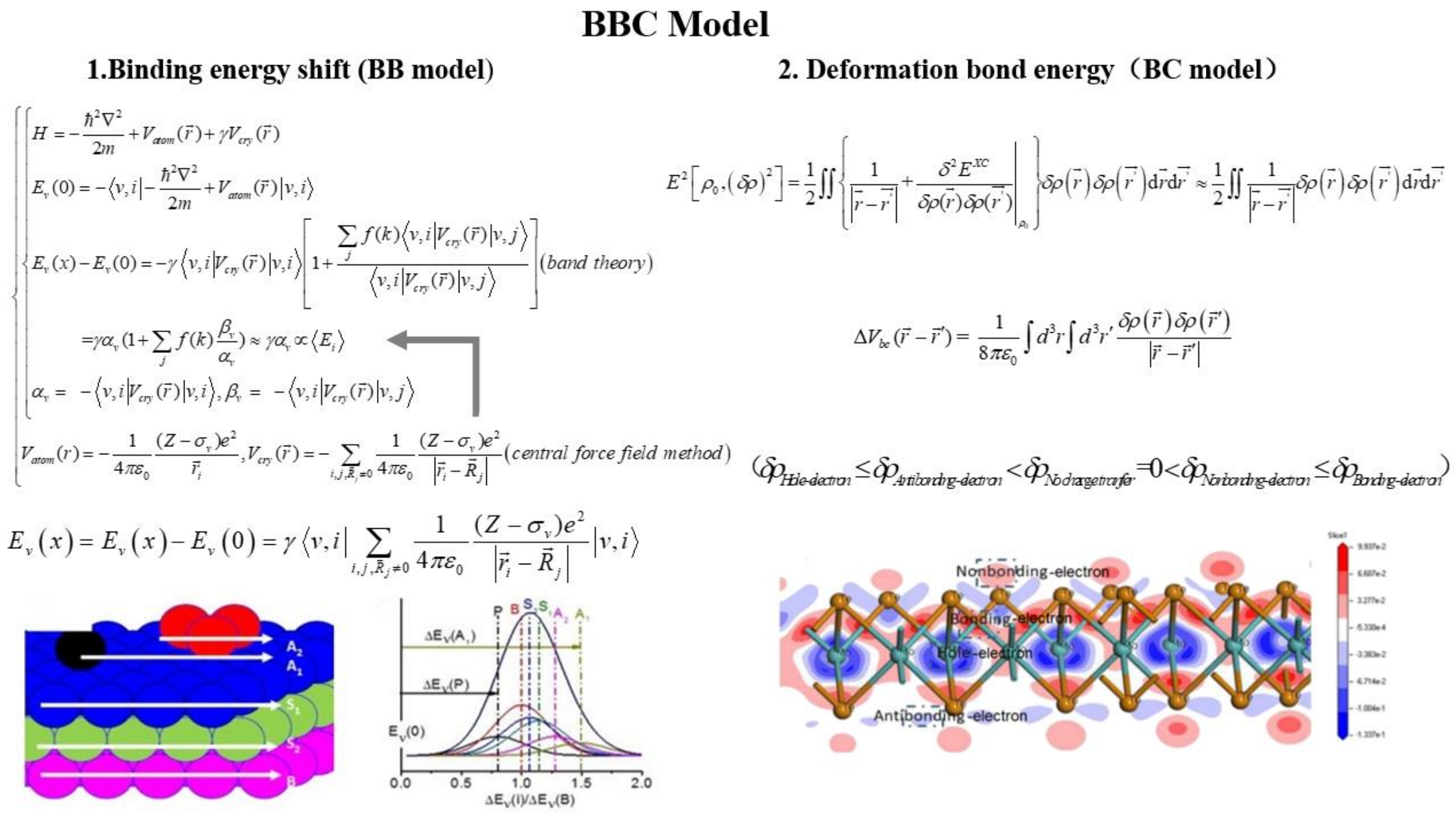

**Fig. S1** Schematic diagram of the BBC model.

The BBC model is quantified chemical bonds by binding energy shift and deformation charge density. For the binding energy (BB) model, we use the central field approximation and Tight-binding (TB) model, which can get the relationship between the energy shift and Hamiltonian. For the bond-charge (BC) model, we use the second-order term of energy expansion and use deformation charge density to calculate bonding states.

For binding energy(BB) model:

$$\begin{cases} H = \int \Psi^{+}(\vec{r})h(\vec{r})\Psi(\vec{r})\mathrm{d}r = \sum_{l,l'} C_l^{+}C_{l'}\int a^{*}\left(\vec{r}-\vec{l}\right)h(\vec{r})a(\vec{r}-l')\mathrm{d}r \\ H = \xi(0)\sum_{l}\hat{n}_l - J\sum_{l}\sum_{\rho} C_l^{+}C_{l+\rho} \\ J = \sum_{l,l'} C_l^{+}C_{l'}\int a^{*}\left(\vec{r}-\vec{l}\right)\left[V(\vec{r})-v_a\left(\vec{r}-\vec{l}\right)\right]a\left(\vec{r}-\vec{l}'\right)\mathrm{d}r \\ \sum_{l}\hat{n}_l = C_l^{+}C_l \end{cases} \quad (1)$$

$V(r)$ is the potential and $v_a\left(\vec{r}-\vec{l}\right)$ is the atomic potential. $\xi(0)$ is the local orbital electron energy. $\hat{n}_l$ represents the electron number operator in the Wanier representation.

$$\begin{cases}\xi(0)=\left(E_n^a-A_n\right)\\ A_n=-\int a_n^*\left(\vec{r}-\vec{l}\right)\left[V\left(\vec{r}\right)-v_a\left(\vec{r}-\vec{l}\right)\right]a_n\left(\vec{r}-\vec{l}\right)\mathrm{d}r\\ E_n^a=\int a_n^*\left(\vec{r}-\vec{l}\right)\left[-\frac{\hbar^2}{2m}\nabla^2+v_a\left(\vec{r}-\vec{l}\right)\right]a_n\left(\vec{r}-\vec{l}\right)\mathrm{d}r\\ \Psi(\vec{r})=\sum_l C_l a(\vec{r}-\vec{l})\end{cases} \tag{2}$$

$E_n^a$ is the atomic energy level, and J is called the overlapping integral. $A_n$ represents the energy level shift $E_n^a$ caused by the potential of (N-1) atoms outside the lattice point *l* in the lattice

Consider the effect of external fields on energy level shifts:

$$\begin{cases}H'=\xi(0)\sum_l\hat{n}_l-J\sum_l\sum_\rho C_l^+C_{l+\rho}=\left(E_n^a-\gamma A_n\right)\sum_l\hat{n}_l-\gamma J\sum_l\sum_\rho C_l^+C_{l+\rho}\\ \Delta E_v(B)=A_n\sum_l\hat{n}_l,\ \ \Delta E_v(x)=\gamma A_n\sum_l\hat{n}_l\\ V_{cry}(\vec{r}_i-\vec{l}_j)=V\left(\vec{r}\right)-v_a\left(\vec{r}-\vec{l}\right)\\ \gamma V_{cry}(\vec{r}_i-\vec{l}_j)=\gamma\sum_{i,j,l_j\neq0}\frac{1}{4\pi\varepsilon_0}\frac{Z'e^2}{\left|\vec{r}_i-\vec{l}_j\right|}=\gamma\sum_{i,j,l_j\neq0}\frac{1}{4\pi\varepsilon_0}\frac{(Z-\sigma_v)e^2}{\left|\vec{r}_i-\vec{l}_j\right|}\\ \Delta E_v(x)=\gamma\int a_n^*\left(\vec{r}-\vec{l}\right)\sum_{i,j,l_j\neq0}\frac{1}{4\pi\varepsilon_0}\frac{(Z-\sigma_v)e^2}{\left|\vec{r}_i-\vec{l}_j\right|}a_n\left(\vec{r}-\vec{l}\right)\mathrm{d}r\sum_l\hat{n}_l\\ \Delta E_v(x)=\gamma\left\langle\Psi_v\left(\vec{r}\right)\right|\sum_{i,j,\vec{R}_j\neq0}\frac{1}{4\pi\varepsilon_0}\frac{(Z-\sigma_v)e^2}{\left|\vec{r}_i-\vec{l}_j\right|}\left|\Psi_v\left(\vec{r}\right)\right\rangle\end{cases} \tag{3}$$

The effective positive charge of the ion is $Z'=Z-\sigma_v$, considering the charge shielding effect $\sigma_v$, where *Z* is the nuclear charge. $\Delta E_v(B)$ is represents the energy shift of an atom in an ideal bulk. $\delta\gamma=\gamma-1$ is relative bond energy ratio and *B* indicates bulk atoms.

For bond-charge (BC) model， we consider the positive charge background (b) and the electron (e) as a system, and write their Hamiltonian sums and their interactions, respectively. In addition to electron kinetic energy, only electrostatic Coulomb interactions are considered:

$$\begin{cases} H = H_b + H_e + H_{eb} \\ H_e = \sum_{i=1}^{N} \frac{P_i^2}{2m} + \frac{1}{2} e_1^2 \sum_{i=1}^{N} \sum_{\substack{j=1 \\ i\neq j}}^{N} \frac{1}{|r_i - r_j|} e^{-\mu|r_i - r_j|} \\ H_b = \frac{1}{2} e_1^2 \int d^3x \int d^3x' \frac{n(x)n(x')}{|x - x'|} e^{-\mu|x-x'|} = \frac{1}{2} e_1^2 \left(\frac{N}{V}\right)^2 \int d^3x 4\pi \int dz \frac{e^{-\mu z}}{z} = 4\pi e_1^2 \frac{N^2}{2V\mu^2} \\ H_{eb} = -e_1^2 \sum_{i=1}^{N} \int d^3x \frac{n(x)}{|x - r_i|} e^{-\mu|x-r_i|} = -e_1^2 \sum_{i=1}^{N} \frac{N}{V} 4\pi \int dz \frac{e^{-\mu z}}{z} = -4\pi e_1^2 \frac{N^2}{V\mu^2} \end{cases} \tag{4}$$

The shielding factor $\mu$ is added to the equation. $r_i$ represents the *ith* electronic position. $x$ represents the background position. $e$ is the basic charge, $e_1 = e/\sqrt{4\pi\varepsilon_0}$ . $e_1 n(x)$ is the charge density at background $x$, and $n(x) = N/V$ is a constant.

$$\begin{cases} H_e = \sum_{i=1}^{N} \frac{P_i^2}{2m} + \frac{1}{2} e_1^2 \sum_{i=1}^{N} \sum_{\substack{j=1 \\ i\neq j}}^{N} \frac{1}{|r_i - r_j|} e^{-\mu|r_i - r_j|} \\ \quad = \sum_{k\sigma} \frac{\hbar^2 k^2}{2m} a_{k\sigma}^{\dagger} a_{k\sigma} + \frac{e_1^2}{2V} \sum_{q}{}^{*} \sum_{k\sigma} \sum_{k'\lambda} \frac{4\pi}{q^2 + \mu^2} a_{k+q,\sigma}^{\dagger} a_{k'-q,\lambda}^{\dagger} a_{k'\lambda} a_{k\sigma} + \frac{e_1^2}{2V} \frac{4\pi}{\mu^2} \left(N^2 - N\right) \\ \sum_{i=1}^{N} \frac{P^2}{2m} = \sum_{l'\sigma'} \sum_{l\sigma} a_{k_{l'}\sigma'}^{\dagger} \left\langle k_{l'}\sigma' \left| \frac{P^2}{2m} \right| k_l \sigma \right\rangle = \sum_{l\sigma} \frac{\hbar^2 k_l^2}{2m} a_{k_l\sigma}^{\dagger} a_{k_l\sigma} \\ \frac{1}{2} e_1^2 \sum_{i=1}^{N} \sum_{\substack{j=1 \\ i\neq j}}^{N} \frac{1}{|r_i - r_j|} e^{-\mu|r_i - r_j|} = \frac{e_1^2}{2V} \sum_{q} \sum_{k'} \sum_{\sigma\lambda} \frac{4\pi}{q^2} a_{k+q,\sigma}^{\dagger} a_{k'-q,\lambda}^{\dagger} a_{k'\lambda} a_{k\sigma} \end{cases} \tag{5}$$

In the formula, the $q$ on the sum sign of "$*$" indicates that the part where q=0 is ignored during the sum. When $V \to \infty, N \to \infty$, while keeping $N/V$ constant, the last term to the right of the equal sign causes the average energy $H_e/N$ of each particle to become:

$$\frac{1}{2}4\pi e_1^2\left(\frac{N}{V}\right)\frac{1}{\mu^2}-\frac{1}{2}4\pi e_1^2\left(\frac{N}{V}\right)\frac{1}{N}\frac{1}{\mu^2}$$

The former term is constant, and the latter term tends to zero. If $\mu\to 0$, the former term becomes a divergent term. However, this term just cancels out the divergent $H_b$ and $H_{eb}$ term. Thus, the Hamiltonian of the system becomes

$$H=\sum_{k\sigma}\frac{\hbar^2k^2}{2m}a_{k\sigma}^{\dagger}a_{k\sigma}+\frac{e_1^2}{2V}\sum_q\sum_{k'}\sum_{\sigma\lambda}\frac{4\pi}{q^2}a_{k+q,\sigma}^{\dagger}a_{k'-q,\lambda}^{\dagger}a_{k'\lambda}a_{k\sigma} \tag{6}$$

Electron interactions expressed using electron density:

$$\begin{cases}\hat{V}_{ee}=\frac{e_1^2}{2V}\sum_k\sum_{k'}\sum_{\sigma\lambda}\frac{4\pi}{q^2}a_{k+q,\sigma}^{\dagger}a_{k'-q,\lambda}^{\dagger}a_{k'\lambda}a_{k\sigma}\\ \quad=\frac{1}{2}\sum_{k_1k_2,k_1'k_2'}\sum_{\sigma_1\sigma_2}\left\langle k_1,k_2\left|v\right|k_1',k_2'\right\rangle C_{k_1\sigma_1}^{+}C_{k_2\sigma_2}^{+}C_{k_2'\sigma_2}C_{k_1'\sigma_1}\\ \quad=\frac{1}{4\pi\varepsilon_0}\times\frac{1}{2\left|\vec{r}-\vec{r}'\right|}\int d^3r\int d^3r'\rho\left(\vec{r}\right)\rho\left(\vec{r}'\right)\\ \rho\left(\vec{r}\right)=a_{k+q,\sigma}^{\dagger}a_{k\sigma}=C_{k_1\sigma_1}^{+}\psi_{k_1}^{*}\left(r\right)C_{k_1\sigma_1}\psi_{k_1'}\left(r\right)\\ \left\langle k_1,k_2\left|v\right|k_1',k_2'\right\rangle=\int\frac{e^2\psi_{k_1}^{*}\left(r\right)\psi_{k_2}^{*}\left(r'\right)\psi_{k_1'}\left(r\right)\psi_{k_2'}\left(r'\right)}{4\pi\varepsilon_0\left|r-r'\right|}\mathrm{d}r\mathrm{d}r'\end{cases} \tag{7}$$

Electron interaction terms for density fluctuations:

$$\begin{cases}V_{ee}'=\frac{e_1^2}{2V}\sum_{\vec{k}}\sum_{\vec{k}'}\sum_{\vec{q}}\sum_{\sigma\lambda}\frac{4\pi}{q^2+\mu^2}a_{\vec{k}+\vec{q},\sigma}^{+}a_{\vec{k}'-\vec{q},\lambda}^{+}a_{\vec{k}'\lambda}a_{\vec{k}\sigma}=\frac{1}{4\pi\varepsilon_0}\frac{1}{2\left|\vec{r}-\vec{r}'\right|}\int d^3r\int d^3r'\rho\left(\vec{r}\right)\rho\left(\vec{r}'\right)e^{-\mu(\vec{r}-\vec{r}')}\\ \delta V_{ee}=V_{ee}'-V_{ee}\\ \quad=\frac{1}{4\pi\varepsilon_0}\frac{1}{2\left|\vec{r}-\vec{r}'\right|}\int d^3r\int d^3r'\delta\rho\left(\vec{r}\right)\delta\rho\left(\vec{r}'\right)\\ \quad=\frac{1}{4\pi\varepsilon_0}\frac{1}{2\left|\vec{r}-\vec{r}'\right|}\int d^3r\int d^3r'\rho\left(\vec{r}\right)\rho\left(\vec{r}'\right)(e^{-\mu(\vec{r}-\vec{r}')}-1)\\ \quad=\frac{1}{4\pi\varepsilon_0}\frac{1}{2\left|\vec{r}-\vec{r}'\right|}\int d^3r\int d^3r'\delta'\rho\left(\vec{r}\right)\delta'\rho\left(\vec{r}'\right)-\frac{1}{4\pi\varepsilon_0}\frac{1}{2\left|\vec{r}-\vec{r}'\right|}\int d^3r\int d^3r'\rho\left(\vec{r}\right)\rho\left(\vec{r}'\right)\\ \Delta V_{bc}(\vec{r}-\vec{r}')=\frac{1}{4\pi\varepsilon_0}\delta V_{ee}=\frac{1}{4\pi\varepsilon_0}\frac{1}{2\left|\vec{r}-\vec{r}'\right|}\int d^3r\int d^3r'\delta\rho\left(\vec{r}\right)\delta\rho\left(\vec{r}'\right)\end{cases}$$

(8)

### 2.2 Electronic properties of metal and semiconductor elements

The energy-orbital distribution is very important for metal and semiconductor physical systems and directly determines their electronic properties. We calculated the partial density of states (PDOS) of the 46 metal and semiconductor elements (**Fig. S3**); their cut-off energies are given in **Table S2**. For metals, electrons do not have zero energy at the Fermi surface ($E_f = 0$). From **Fig. S3**, we can see that the *d*-orbital PDOSs of Ca, Sc, Ti, V, Cr, Fe, Co, Ni, Sr, Y, Zr, Nb, Mo, Tc, Ru, Rh, Pd, Ba, Hf, Ta, W, Re, Os, Ir, and Pt had obvious fluctuations and spikes near the Fermi surface. This shows that the *d* electrons were relatively localized, and the main contribution to the structural energy was from the d orbital. The IA and IIA elements Cu, Zn, Ag, Cd, and Au had small fluctuations near the Fermi surface and large ones on its left side; they were relatively smooth near the Fermi level, with comparatively strong non-local properties. The *d* orbitals of Li, Be, C, Na, Mg, Al, Si, K, Ge, Rb, In, Sn, Cs, Lu, Tl, and Pb contributed very little. In addition, the elements of C, Si, and Ge with diamond structure had semiconductor properties; in them, there was no electron distribution near the Fermi level, and there were obvious band gaps. From **Fig. S4**, the band gaps of C, Si, and Ge were 3.988 eV, 0.591 eV, and 0.494 eV, respectively.

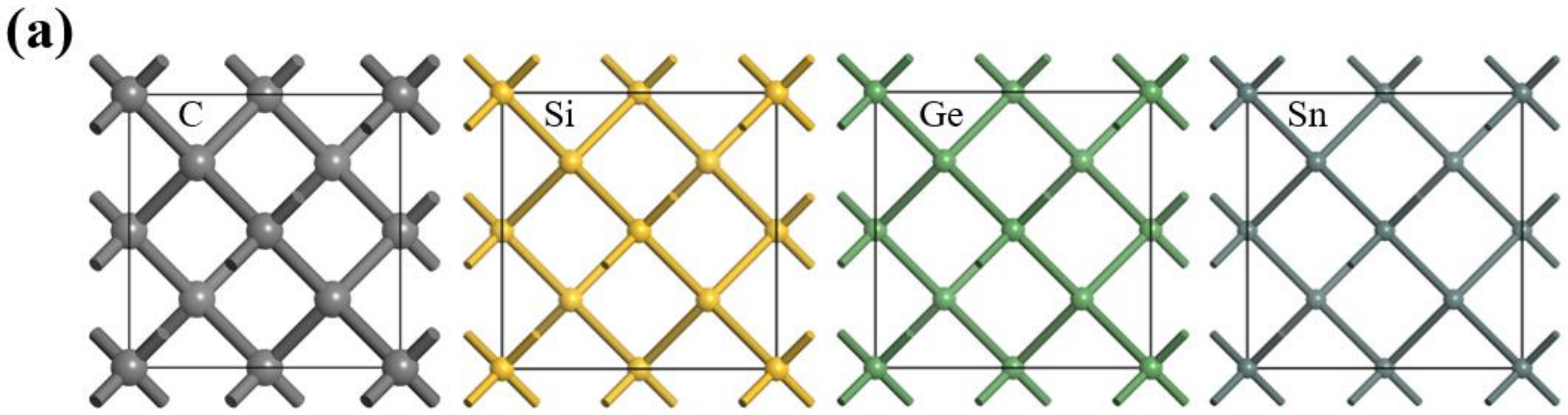

(b)

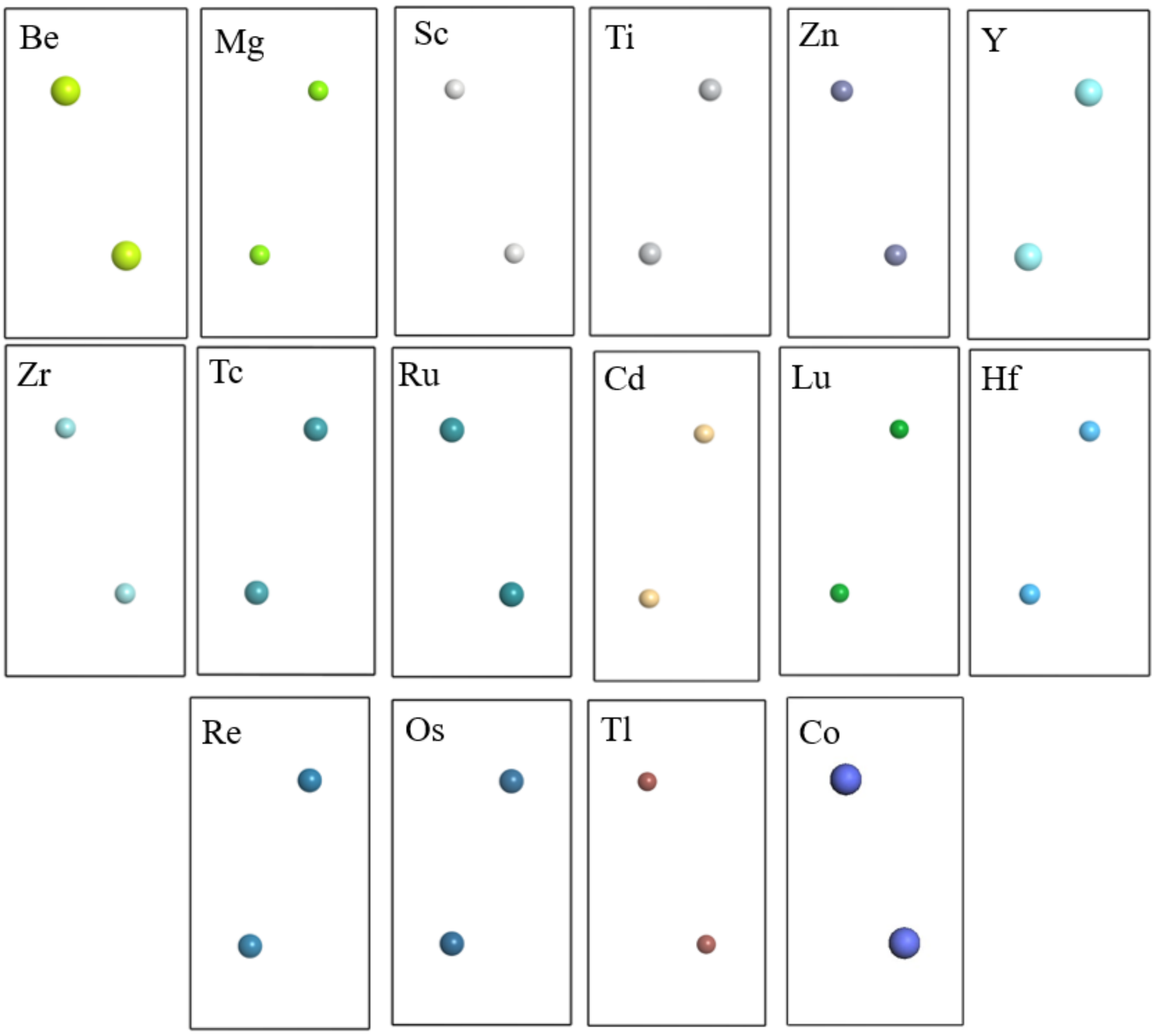

(c)

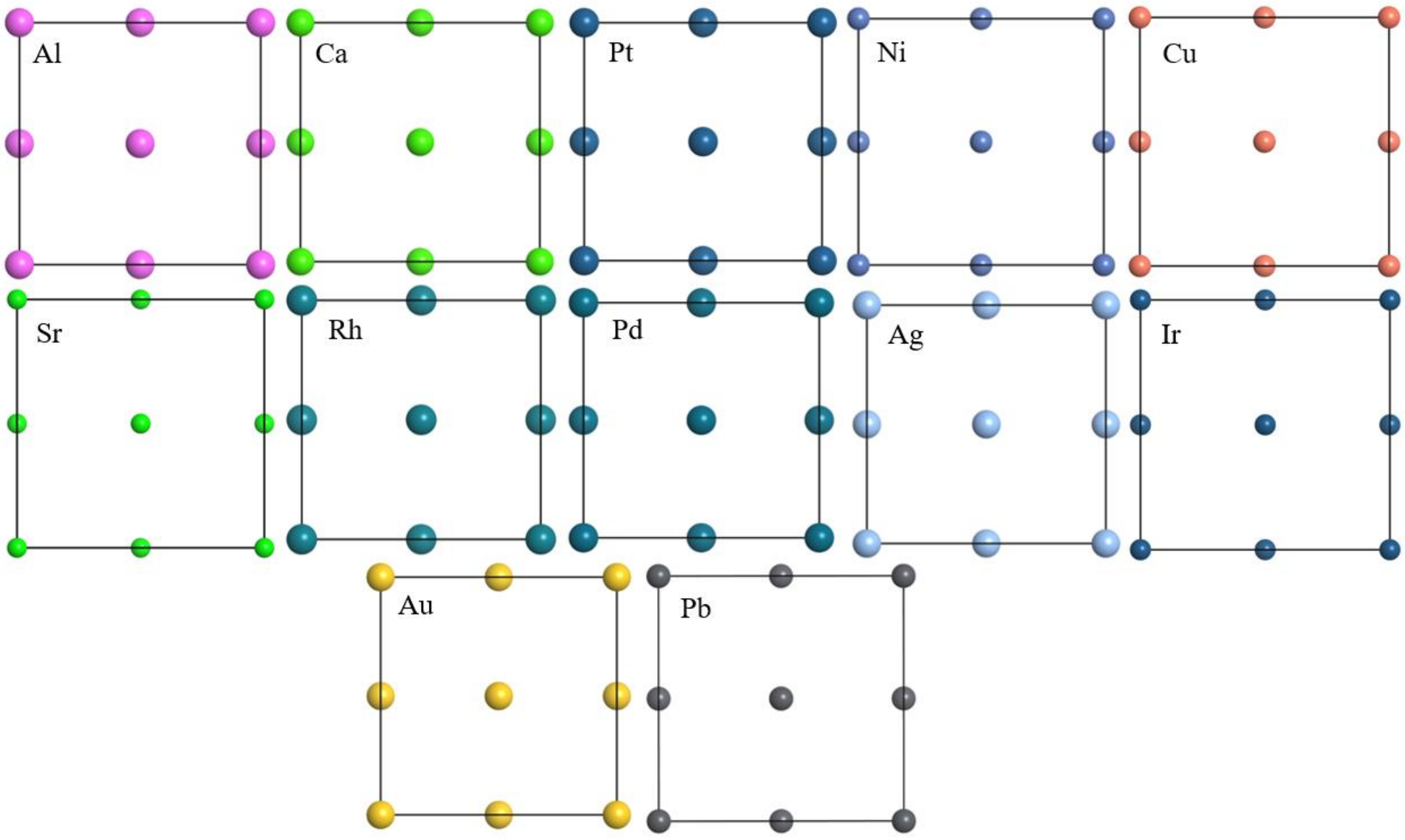

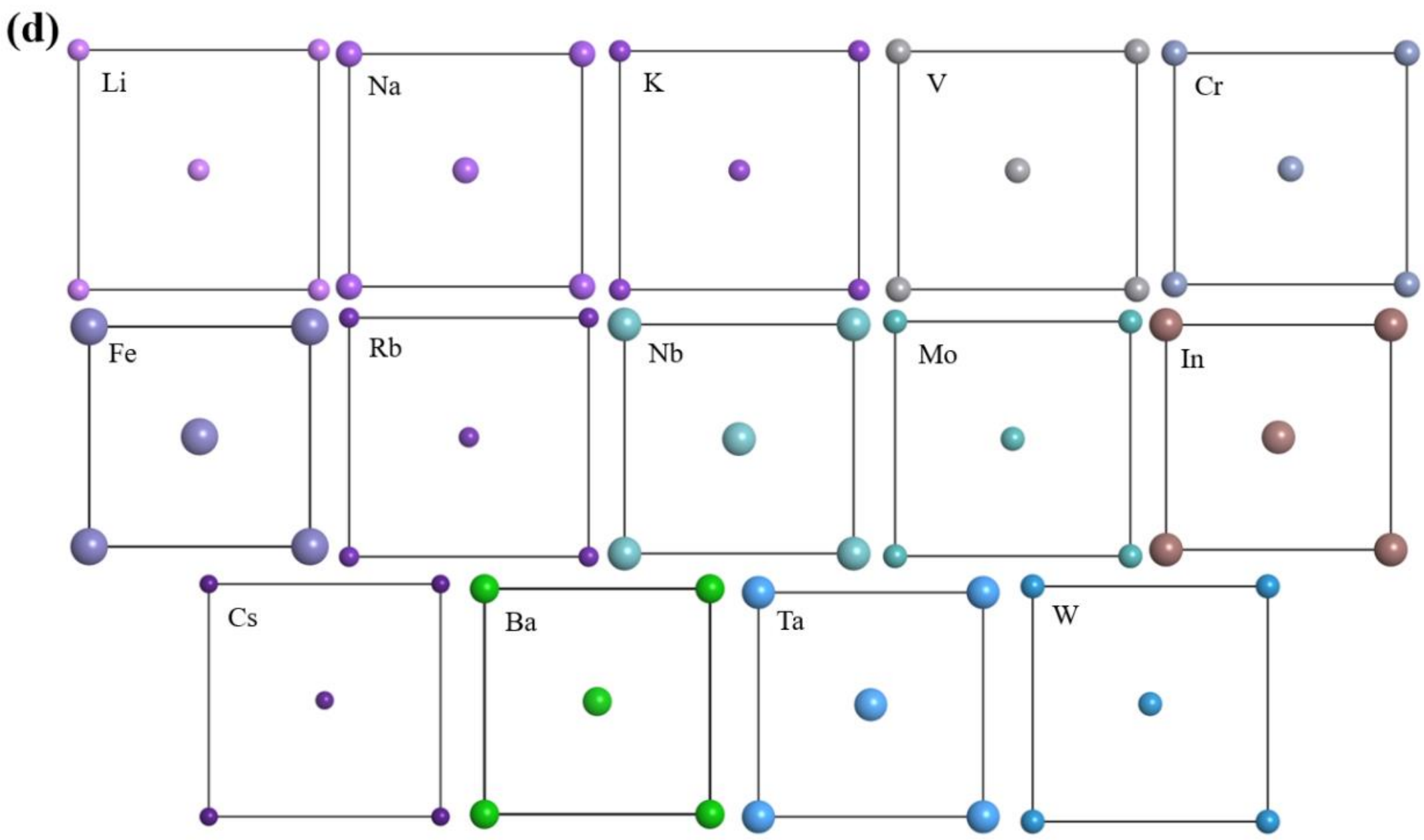

**Fig. S2** Initial geometric structures of 46 metal and semiconductor elements: (a) diamond structure; (b) hcp structure; (c) fcc structure; (d) bcc structure.

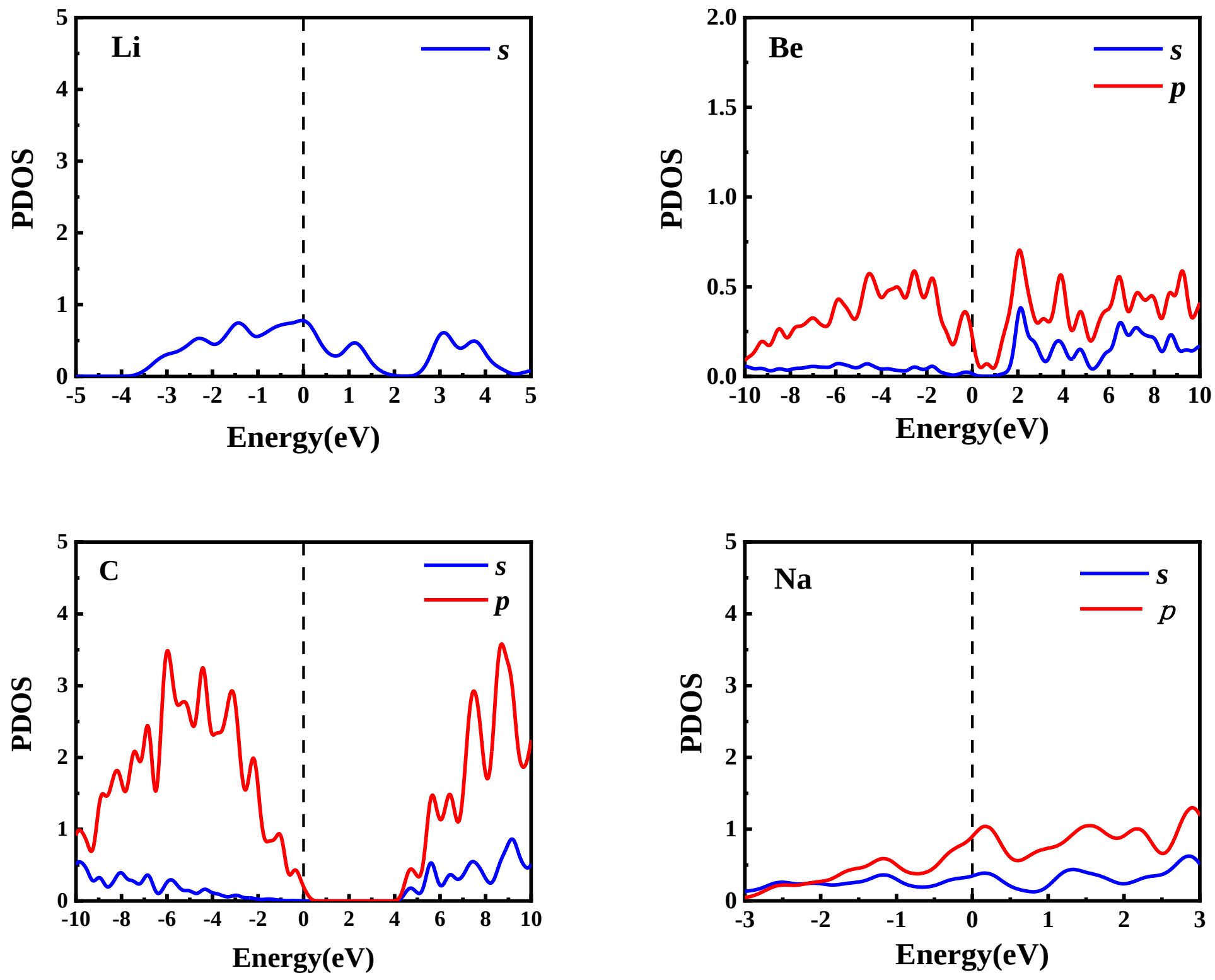

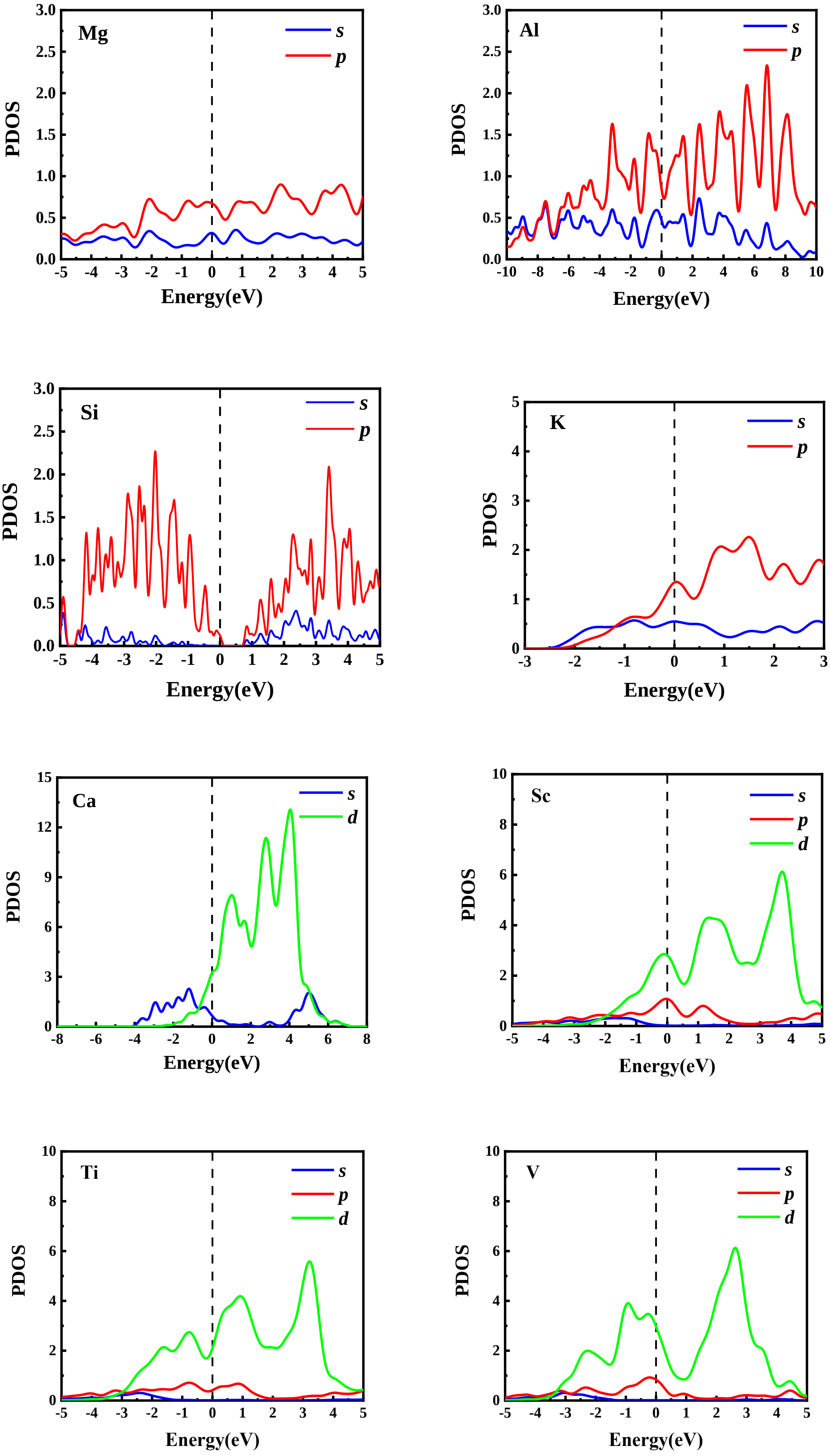

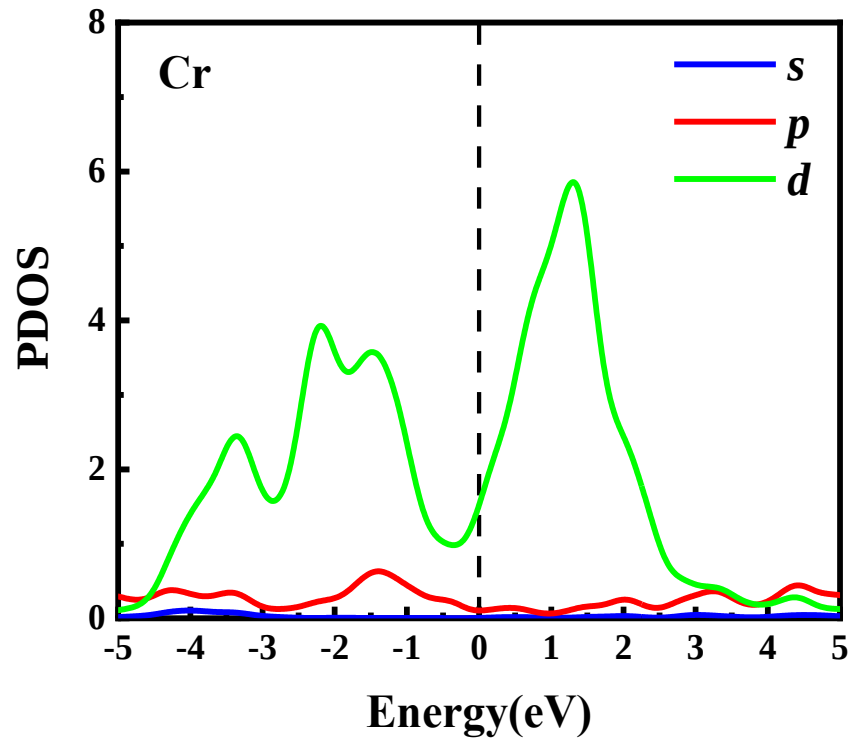
Cr
s
p
d
PDOS
Energy(eV)
0 2 4 6 8
-5 -4 -3 -2 -1 0 1 2 3 4 5

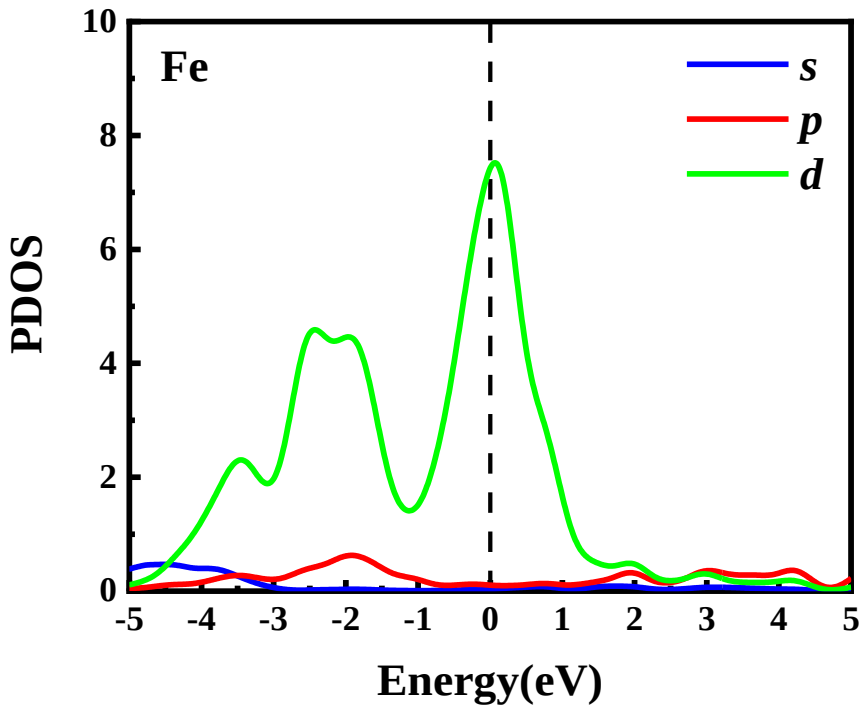
Fe
s
p
d
PDOS
Energy(eV)
0 2 4 6 8 10
-5 -4 -3 -2 -1 0 1 2 3 4 5

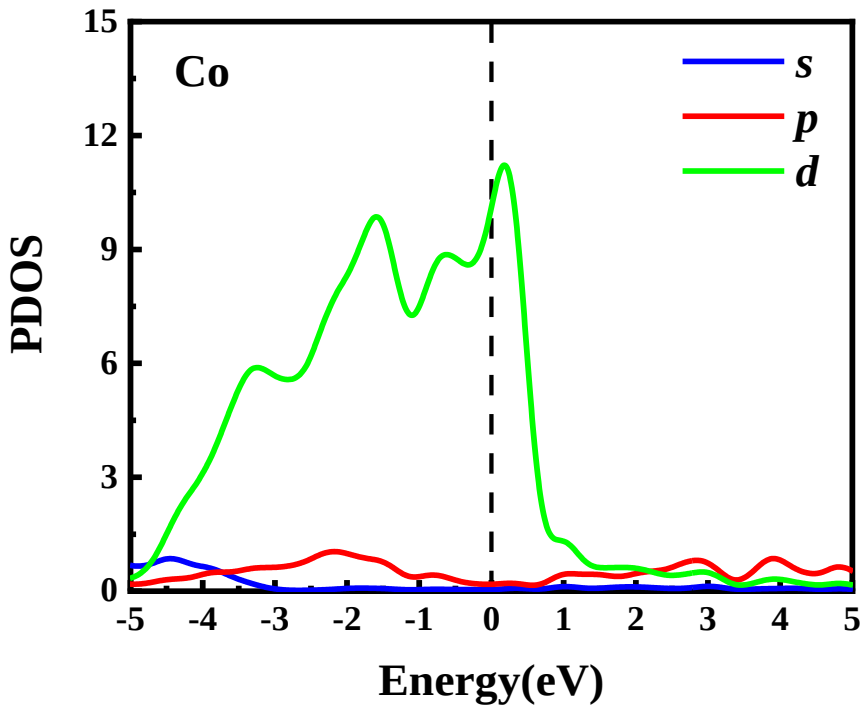
Co
s
p
d
PDOS
Energy(eV)
0 3 6 9 12 15
-5 -4 -3 -2 -1 0 1 2 3 4 5

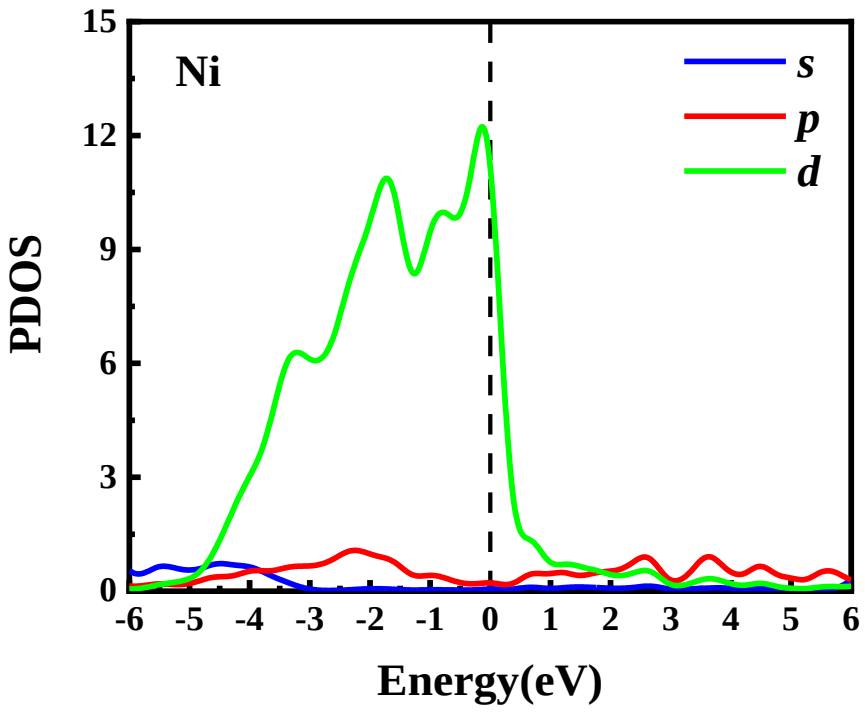
Ni
s
p
d
PDOS
Energy(eV)
0 3 6 9 12 15
-6 -5 -4 -3 -2 -1 0 1 2 3 4 5 6

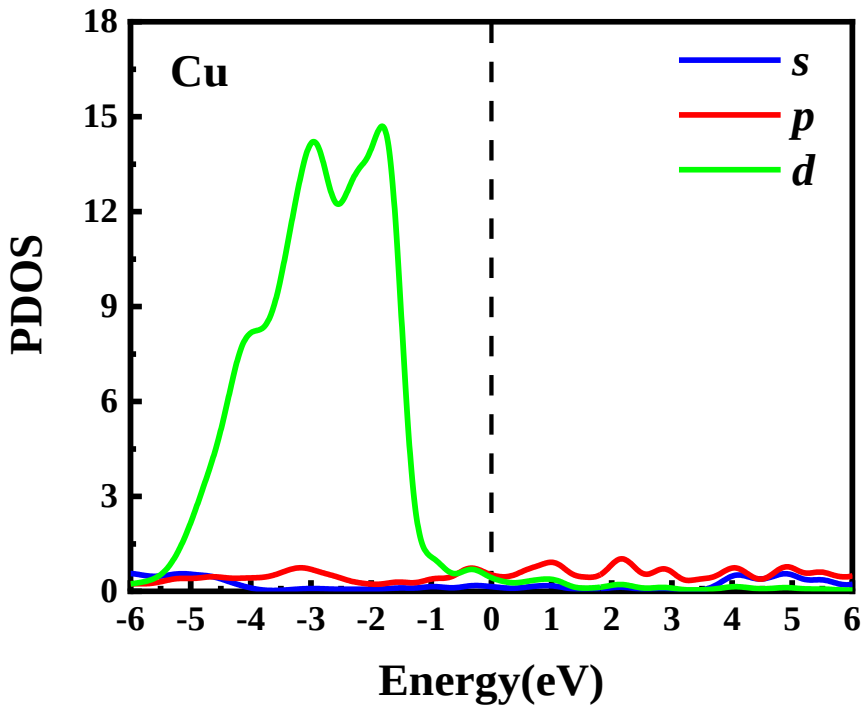
Cu
s
p
d
PDOS
Energy(eV)
0 3 6 9 12 15 18
-6 -5 -4 -3 -2 -1 0 1 2 3 4 5 6

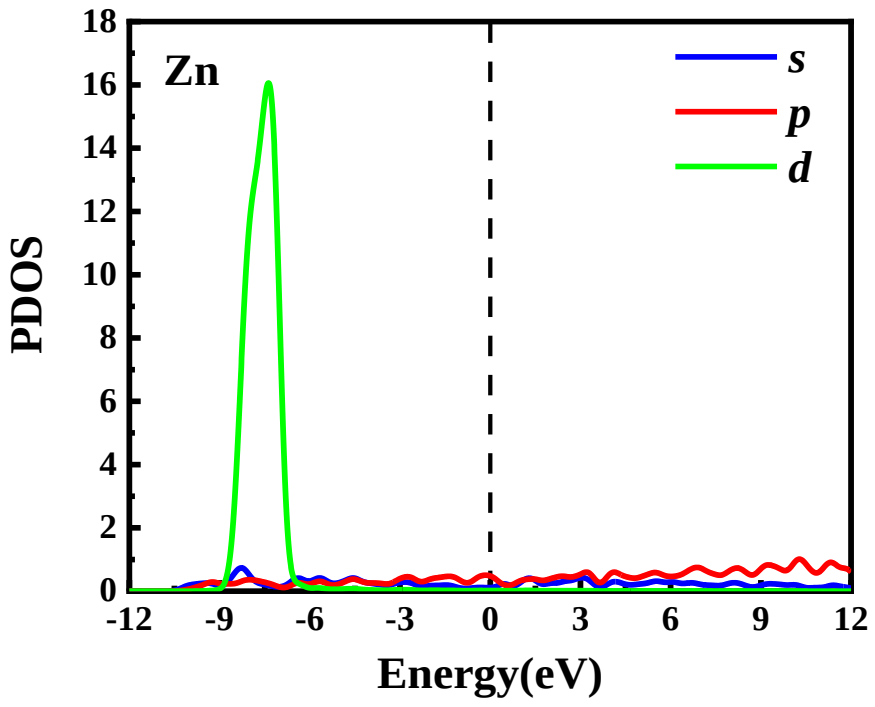
Zn
s
p
d
PDOS
Energy(eV)
0 2 4 6 8 10 12 14 16 18
-12 -9 -6 -3 0 3 6 9 12

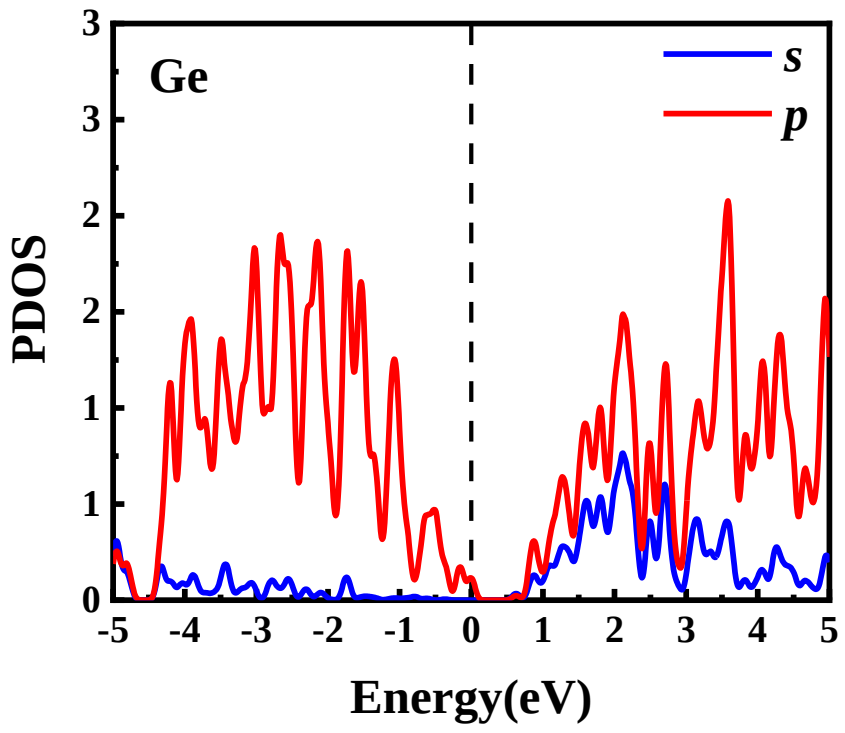
Ge
s
p
PDOS
Energy(eV)
0 1 1 2 2 3 3
-5 -4 -3 -2 -1 0 1 2 3 4 5

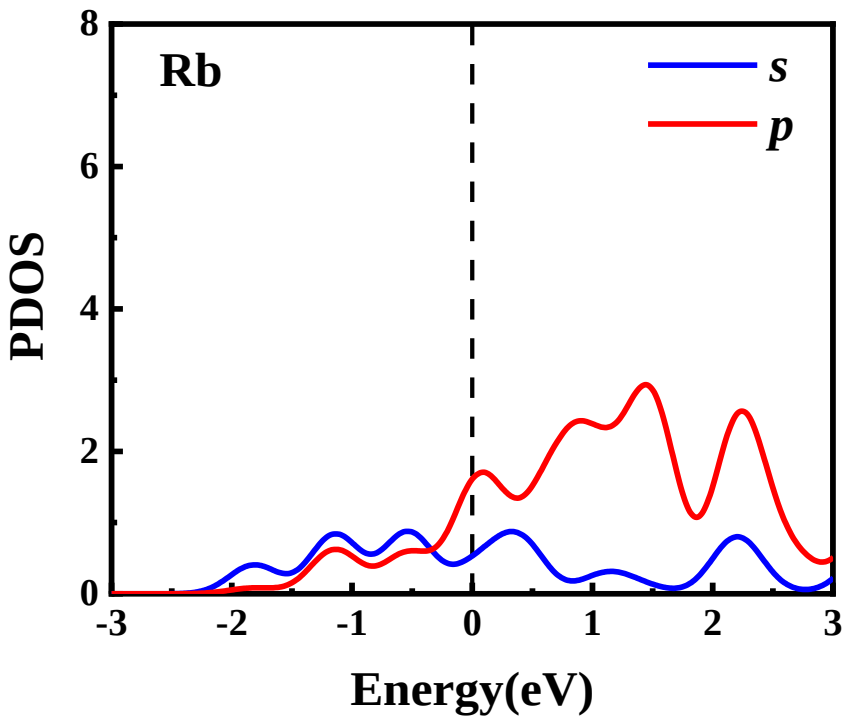
Rb
s
p
PDOS
Energy(eV)
0 2 4 6 8
-3 -2 -1 0 1 2 3

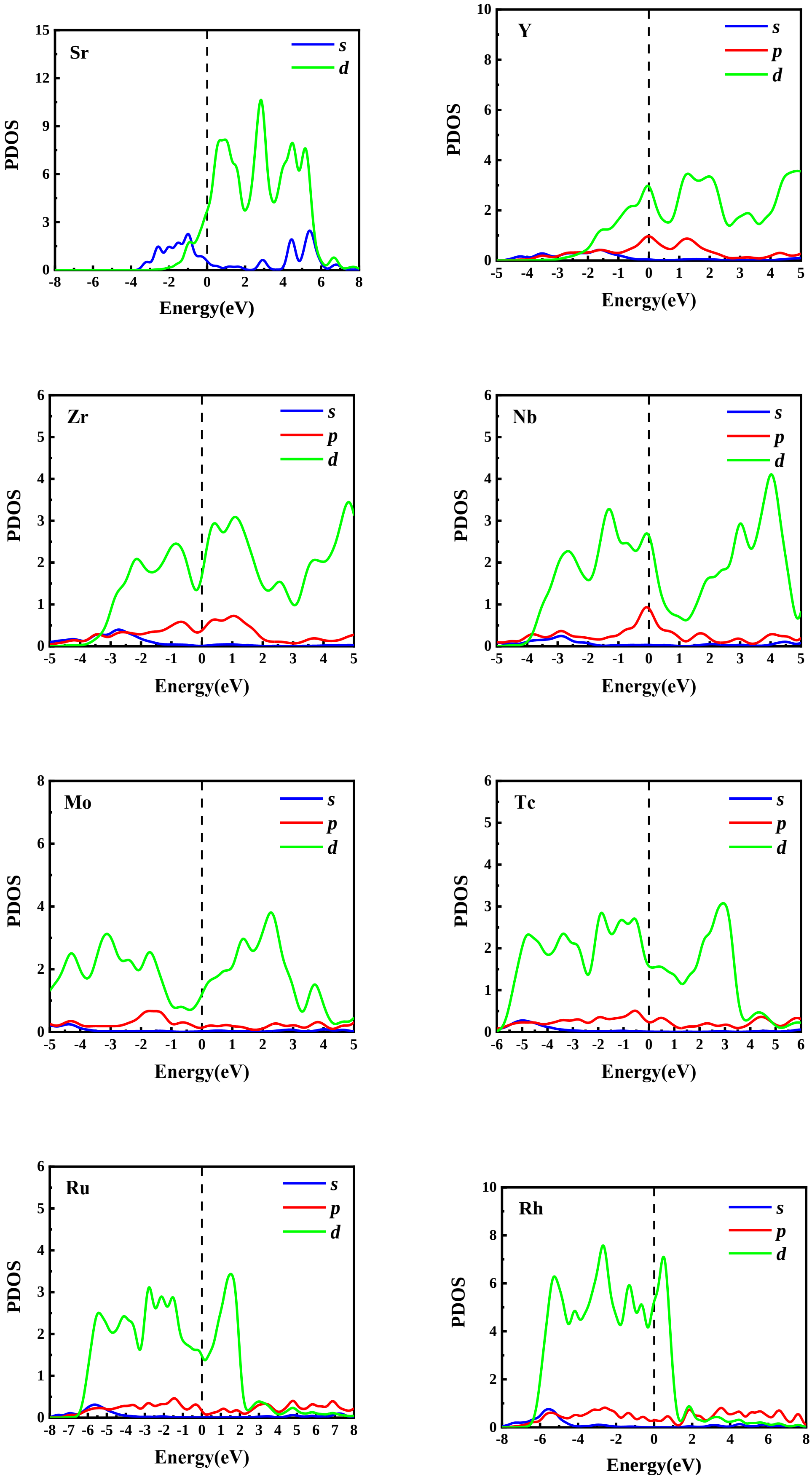
Sr
s
d
PDOS
0 3 6 9 12 15
-8 -6 -4 -2 0 2 4 6 8
Energy(eV)
Y
s
p
d
PDOS
0 2 4 6 8 10
-5 -4 -3 -2 -1 0 1 2 3 4 5
Energy(eV)
Zr
s
p
d
PDOS
0 1 2 3 4 5 6
-5 -4 -3 -2 -1 0 1 2 3 4 5
Energy(eV)
Nb
s
p
d
PDOS
0 1 2 3 4 5 6
-5 -4 -3 -2 -1 0 1 2 3 4 5
Energy(eV)
Mo
s
p
d
PDOS
0 2 4 6 8
-5 -4 -3 -2 -1 0 1 2 3 4 5
Energy(eV)
Tc
s
p
d
PDOS
0 1 2 3 4 5 6
-6 -5 -4 -3 -2 -1 0 1 2 3 4 5 6
Energy(eV)
Ru
s
p
d
PDOS
0 1 2 3 4 5 6
-8 -7 -6 -5 -4 -3 -2 -1 0 1 2 3 4 5 6 7 8
Energy(eV)
Rh
s
p
d
PDOS
0 2 4 6 8 10
-8 -6 -4 -2 0 2 4 6 8
Energy(eV)

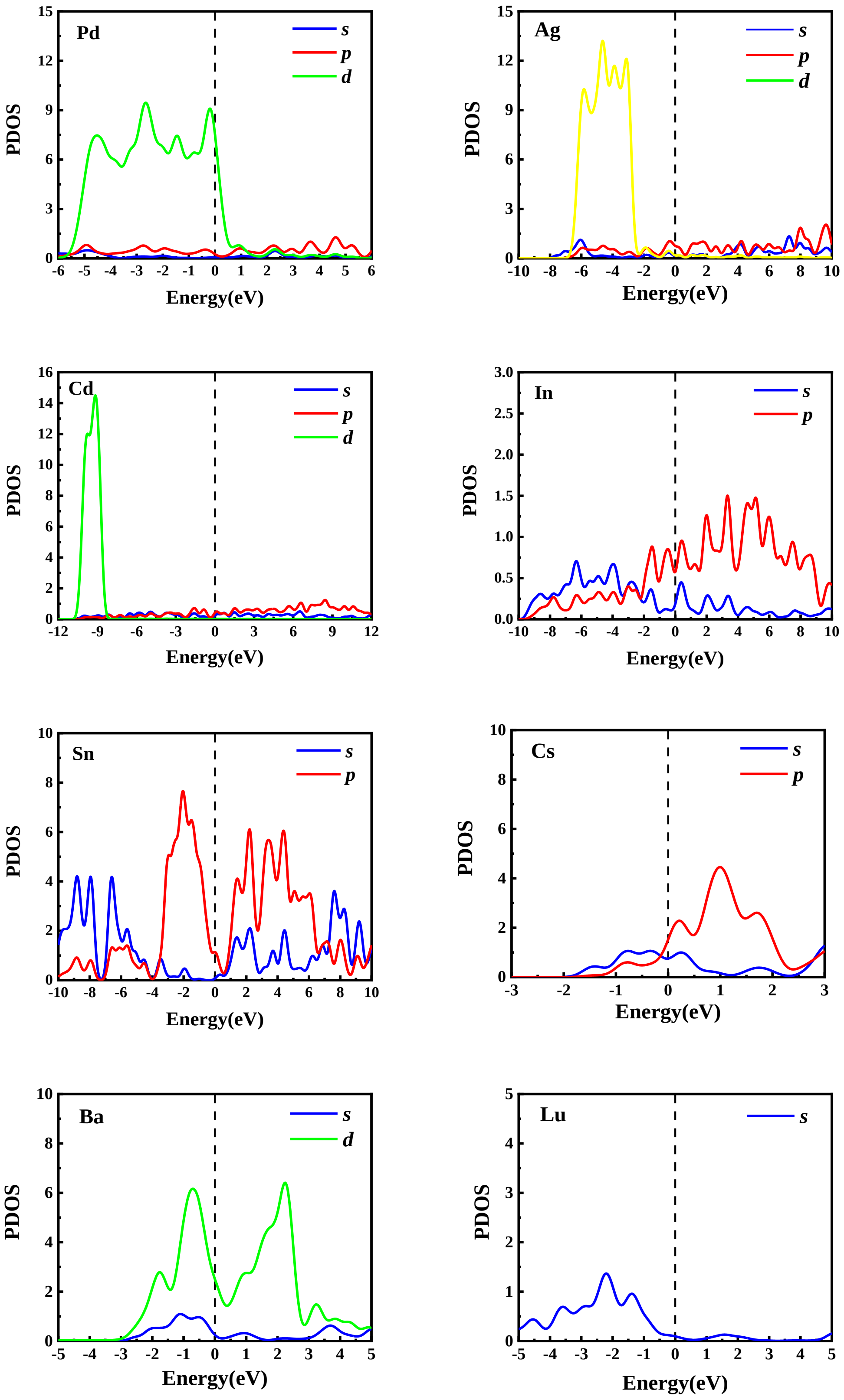

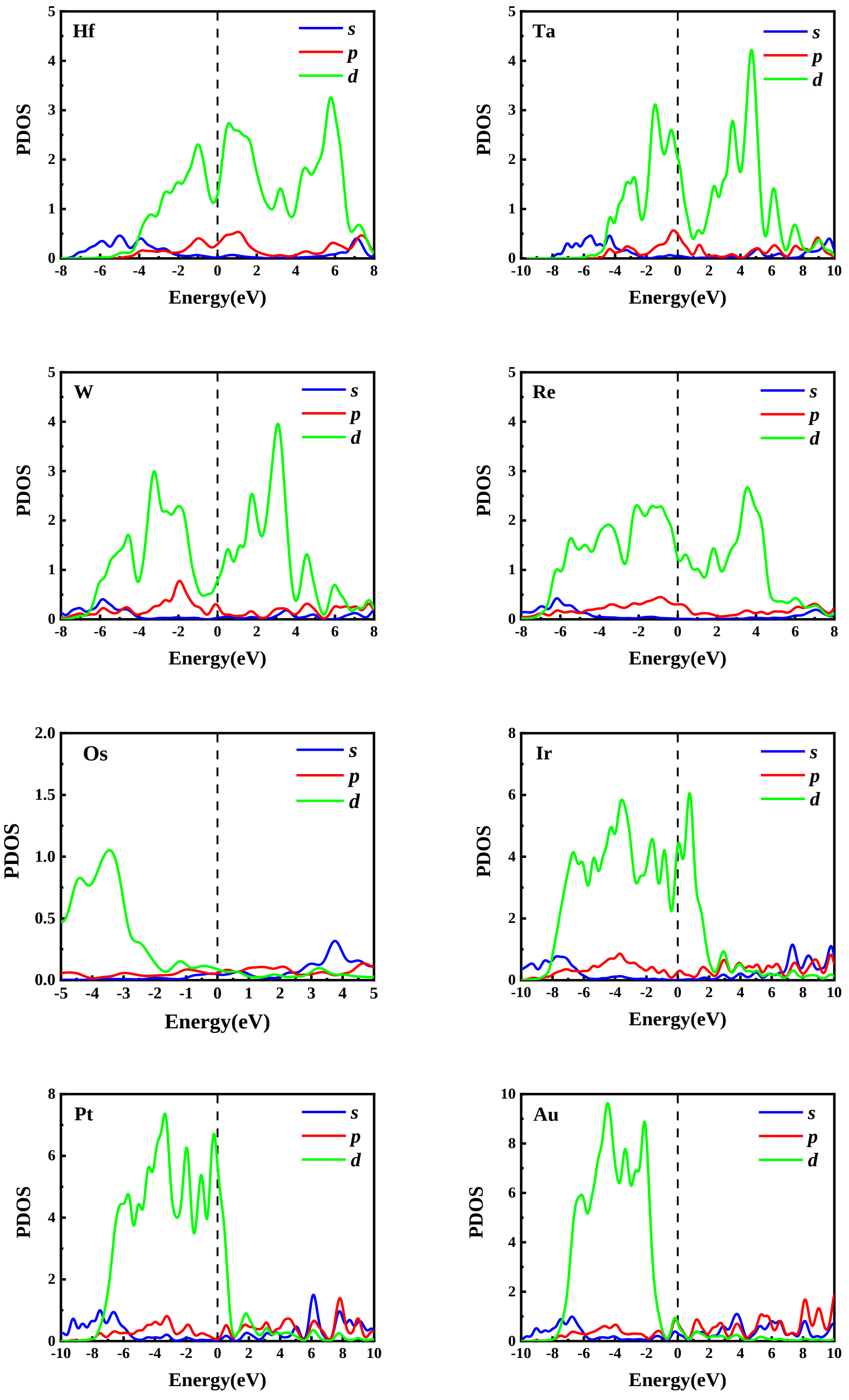

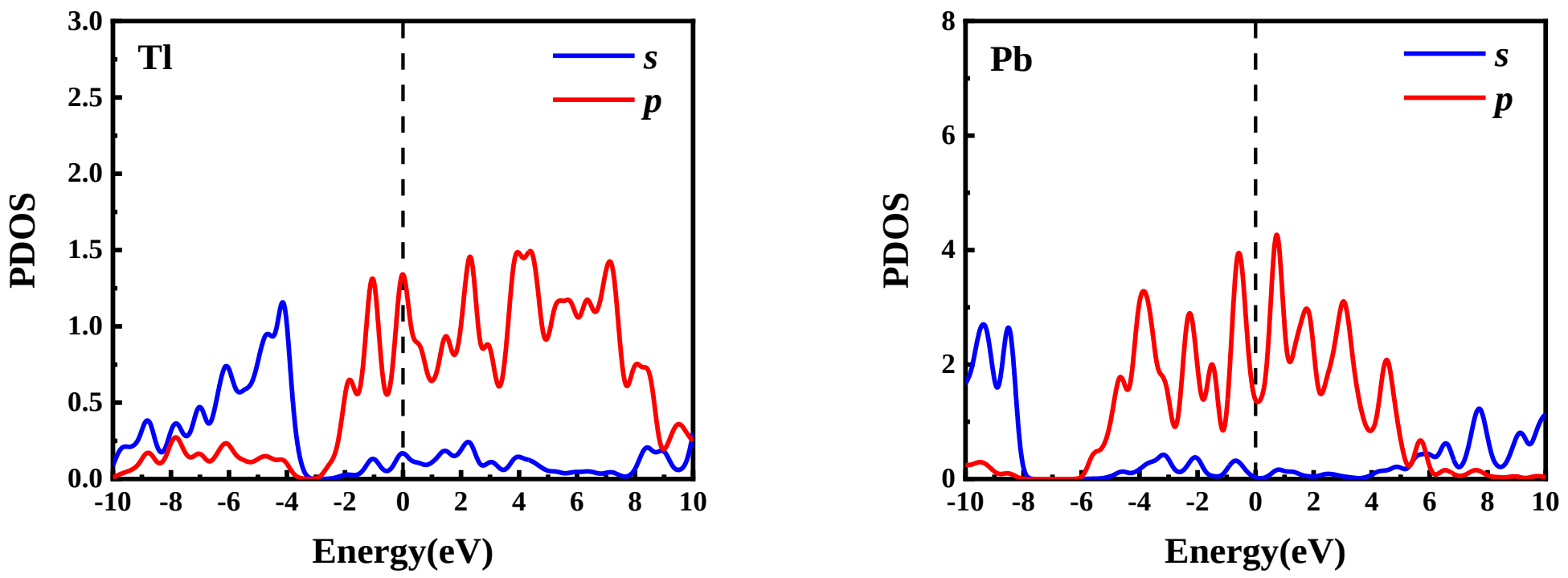

**Fig. S3** Partial density of states (PDOS) of the *s*-, *p*- and *d*- orbitals of 46 metal and semiconductor elements.

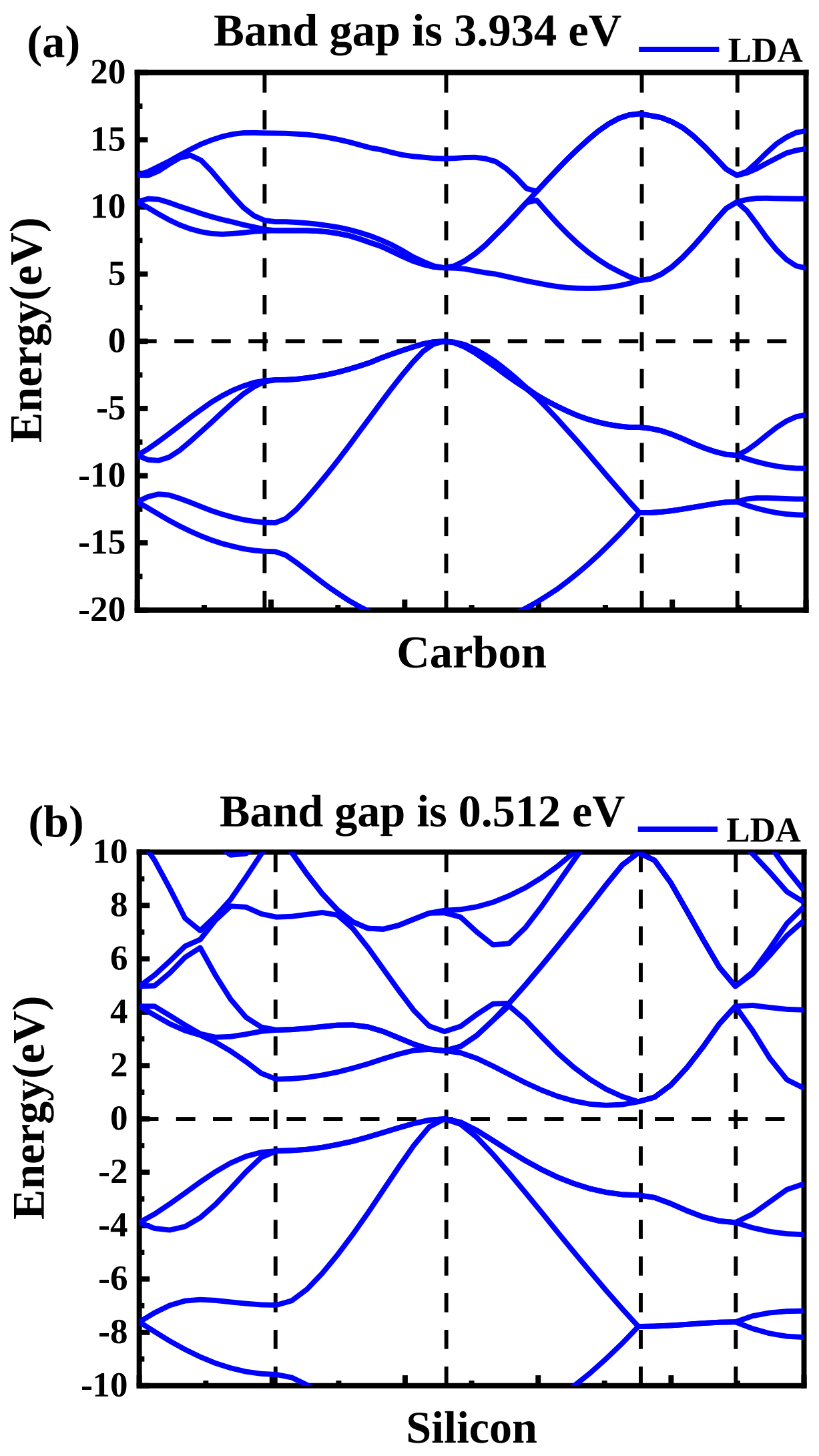

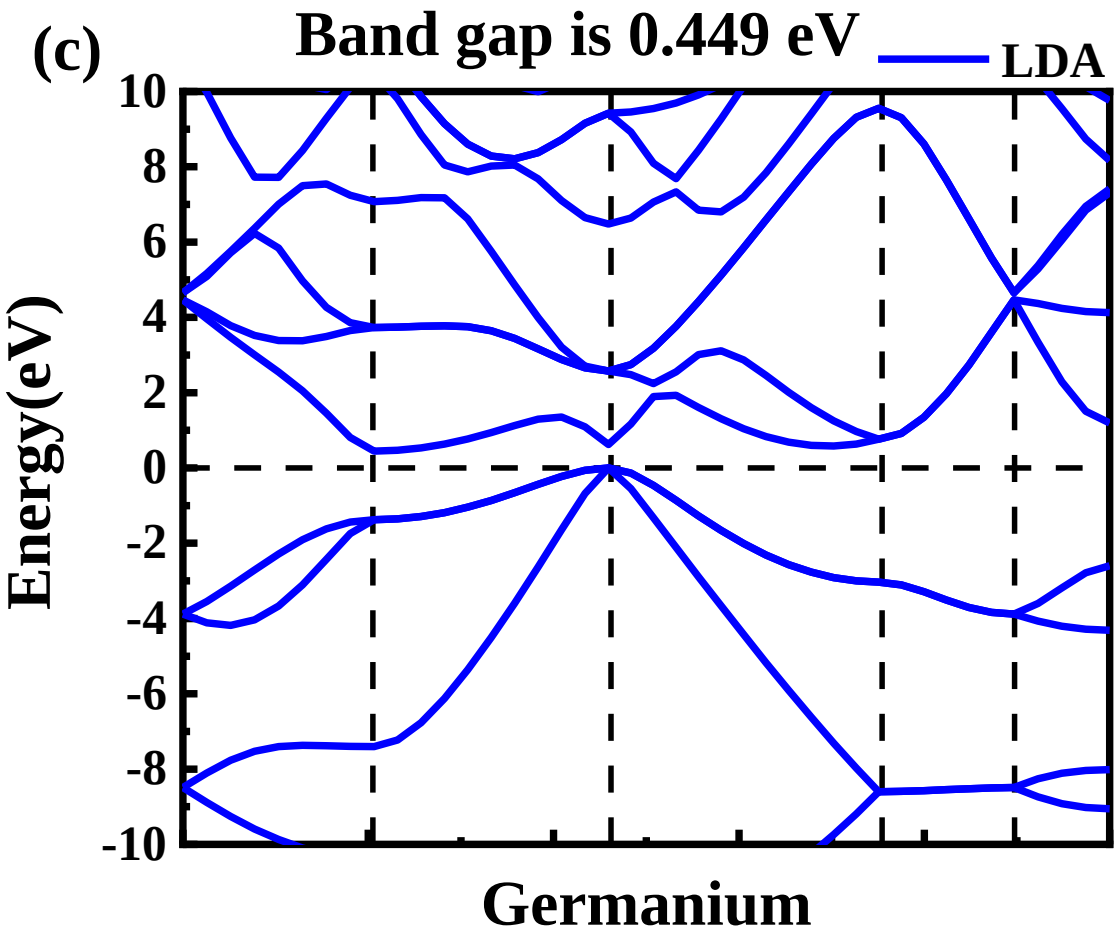

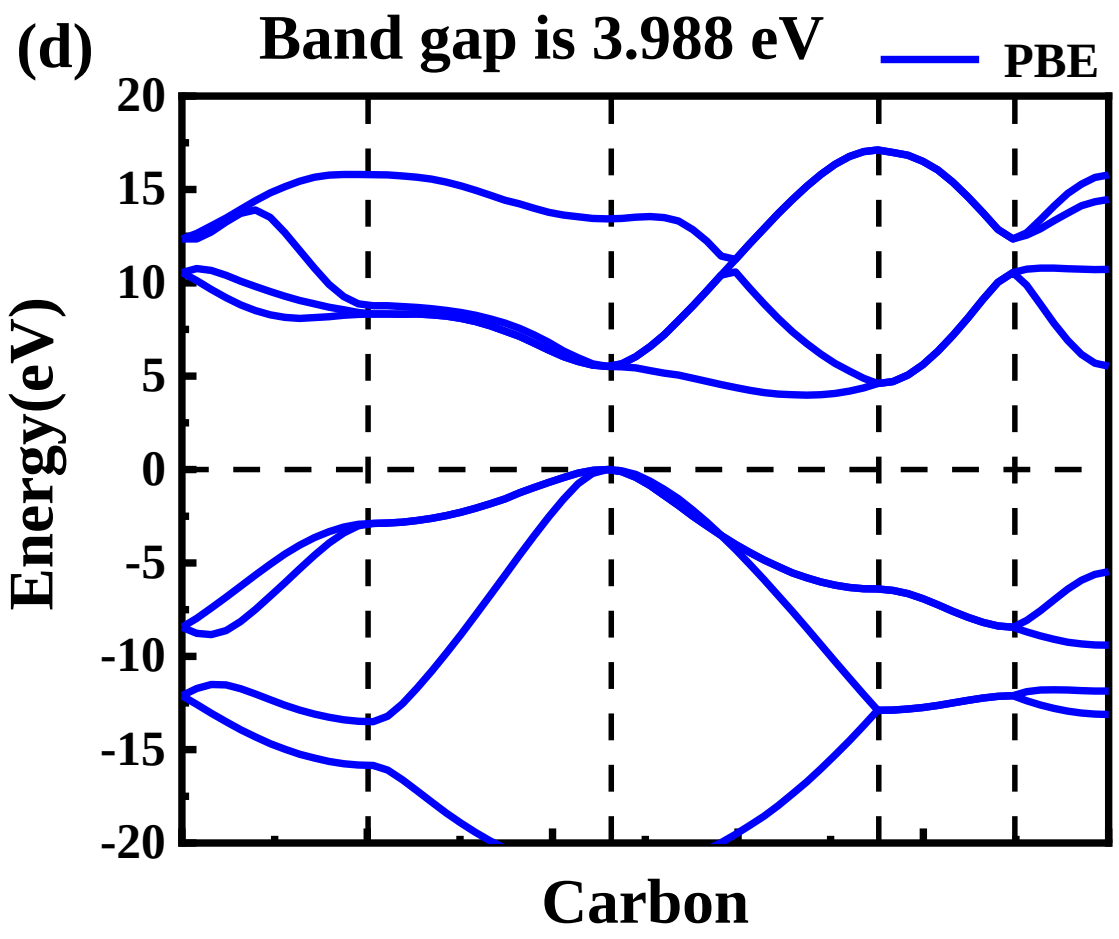

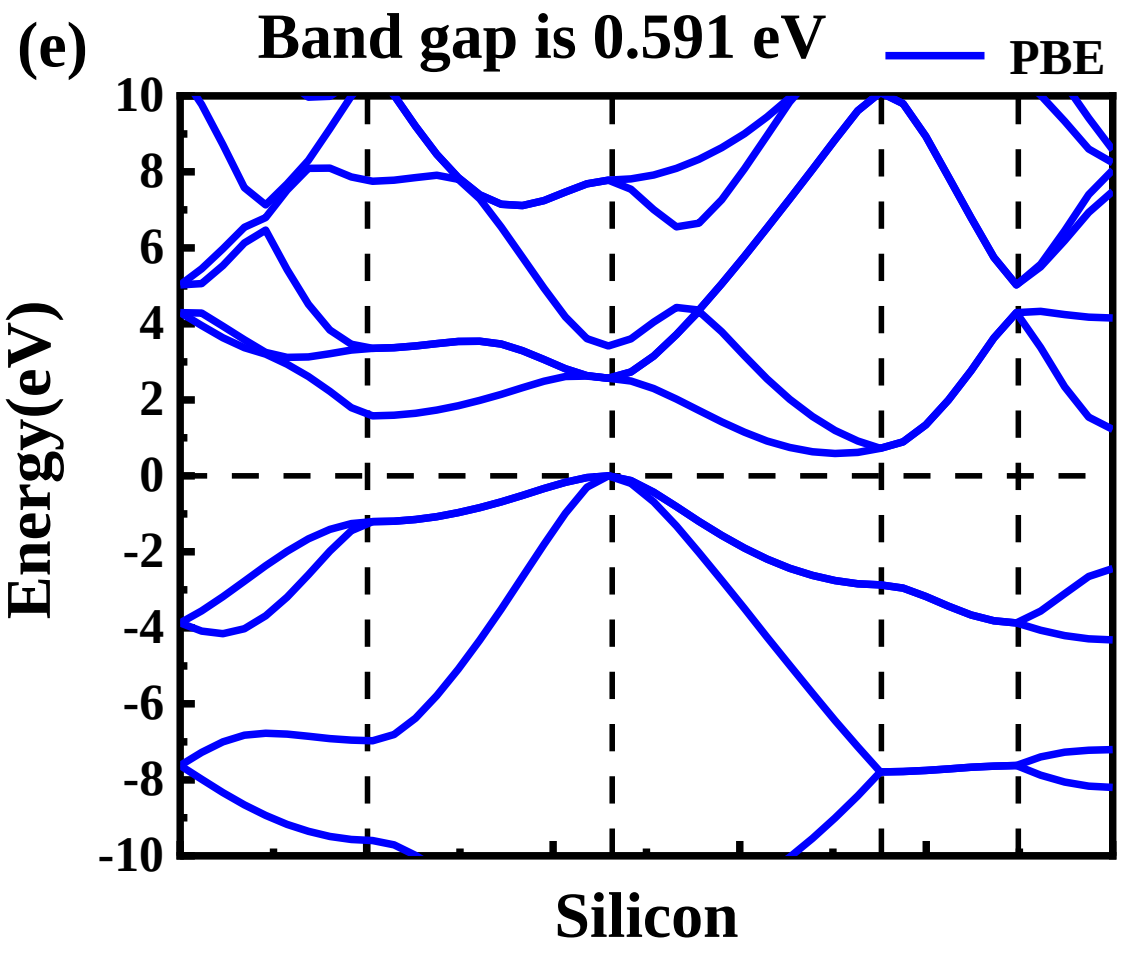

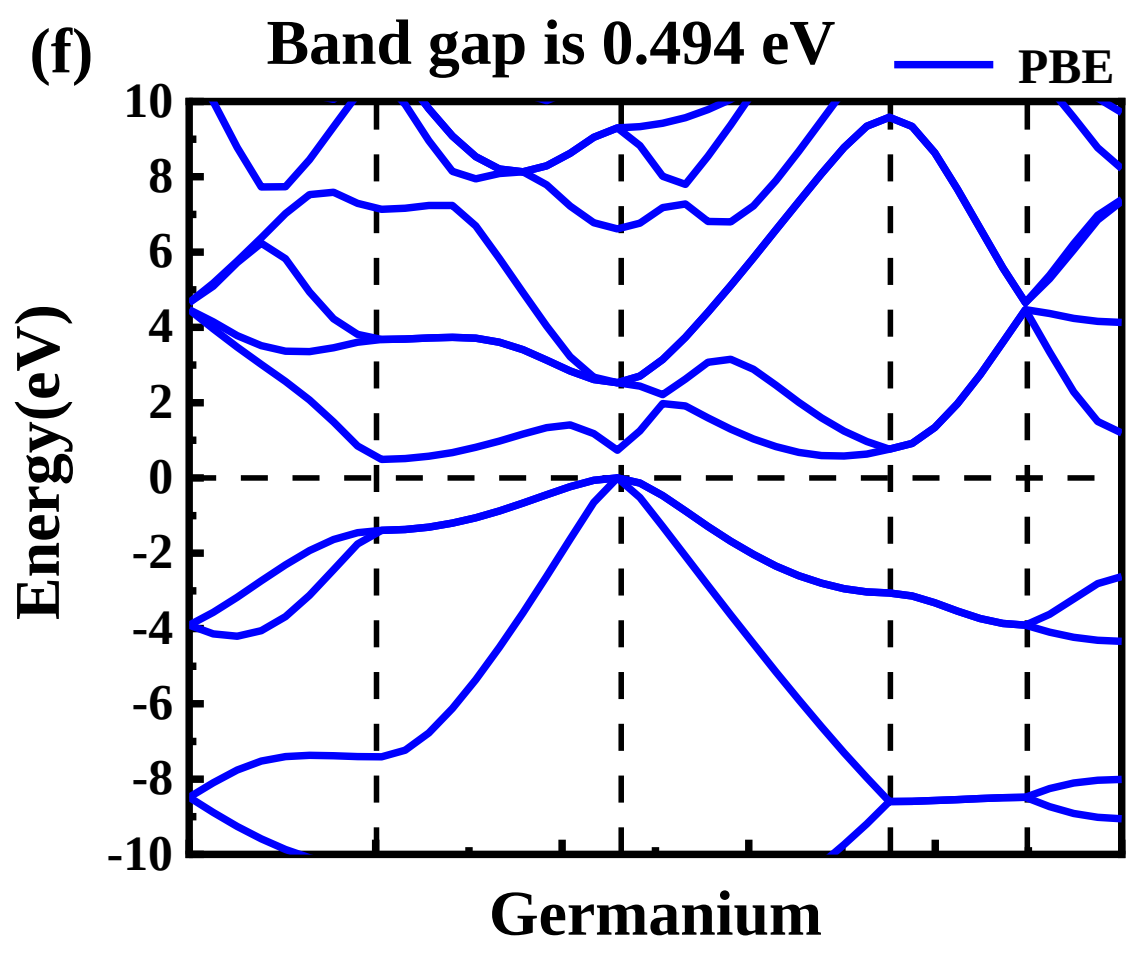

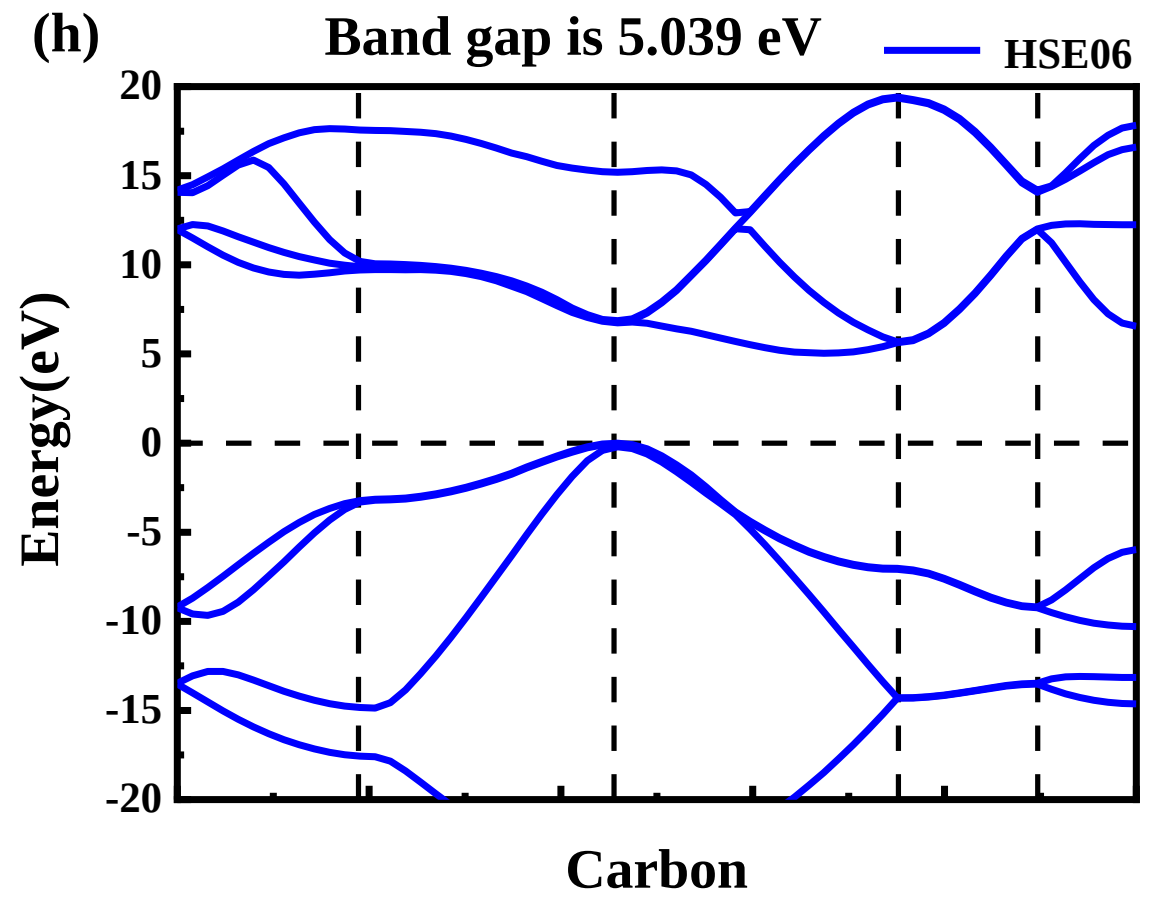

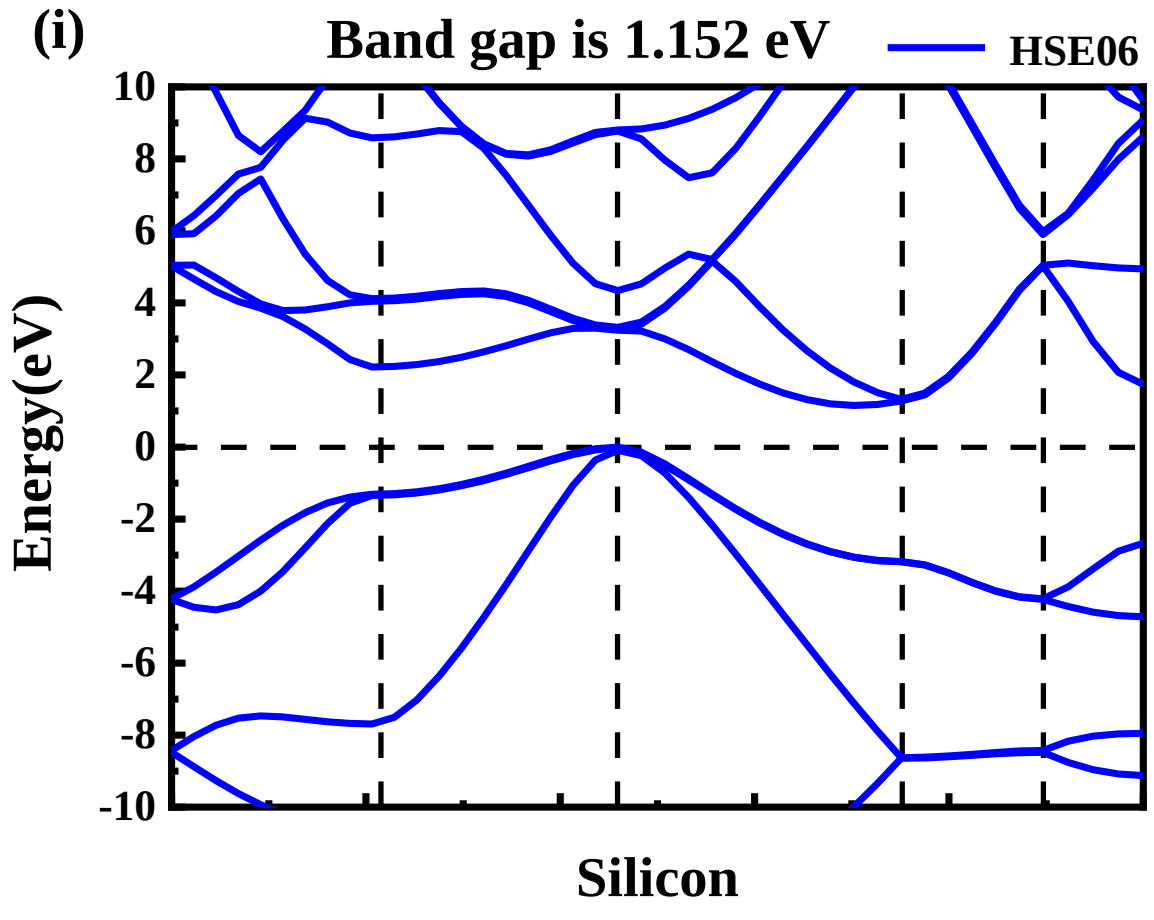

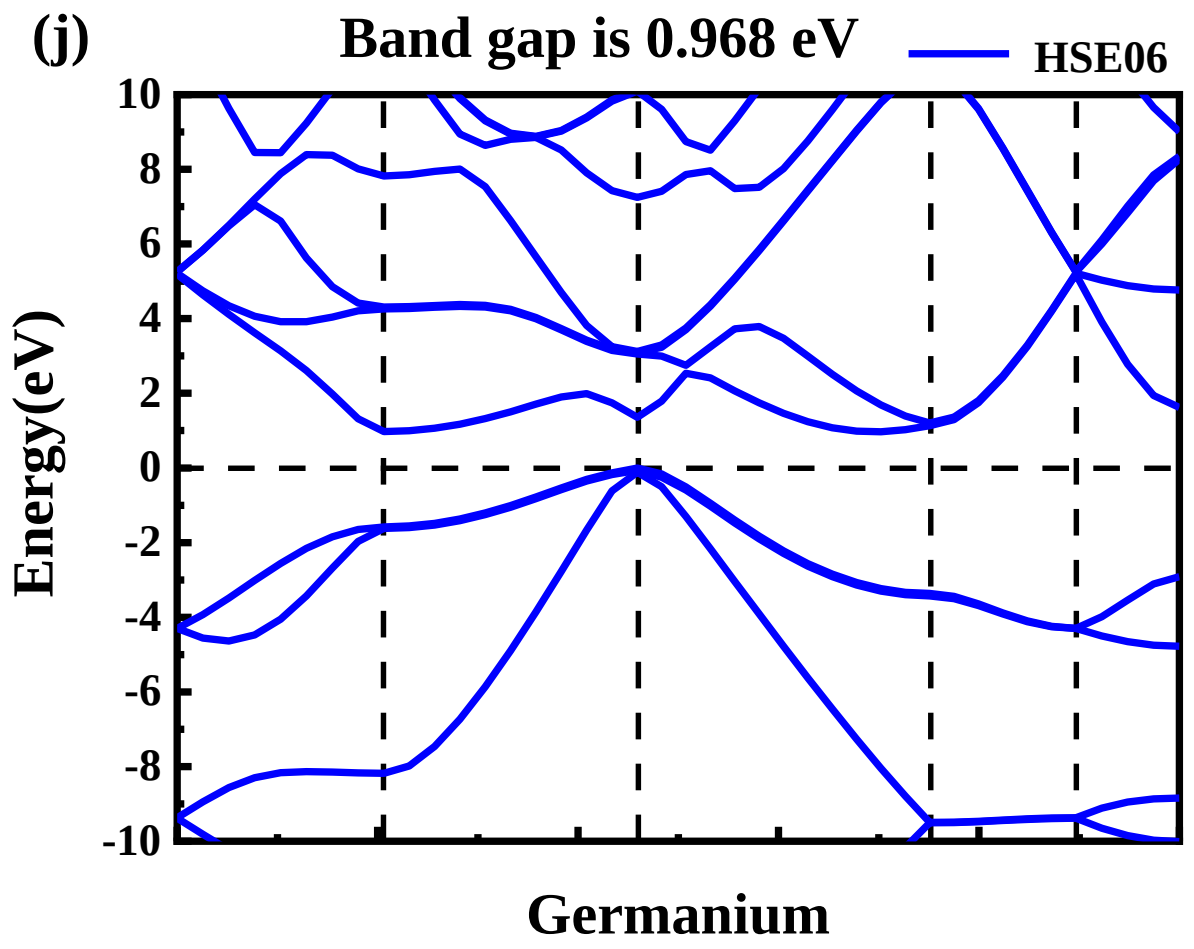

**Fig. S4** Energy band of C, Si, Ge calculated by LDA, PBE and HSE06 functional.

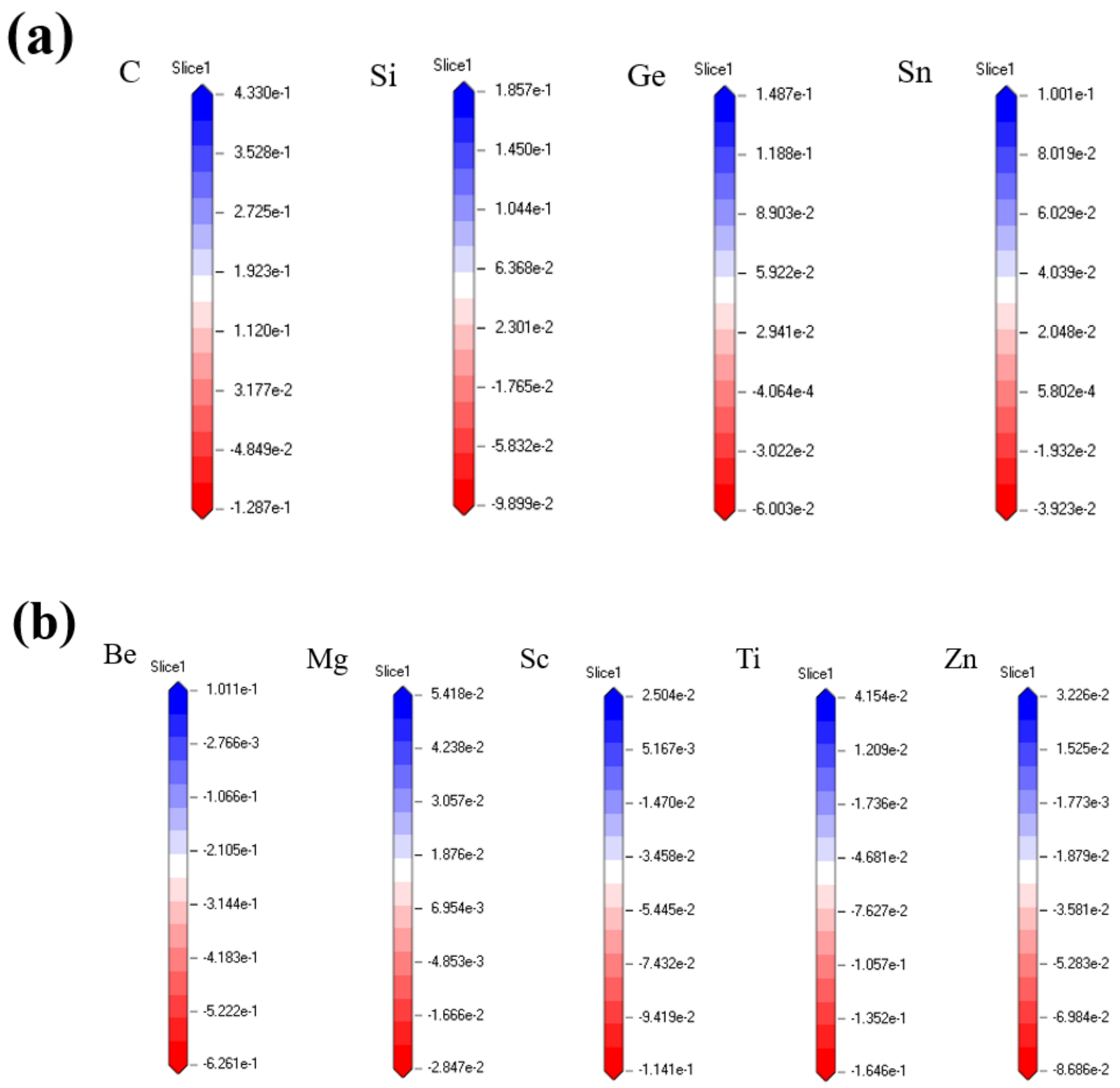

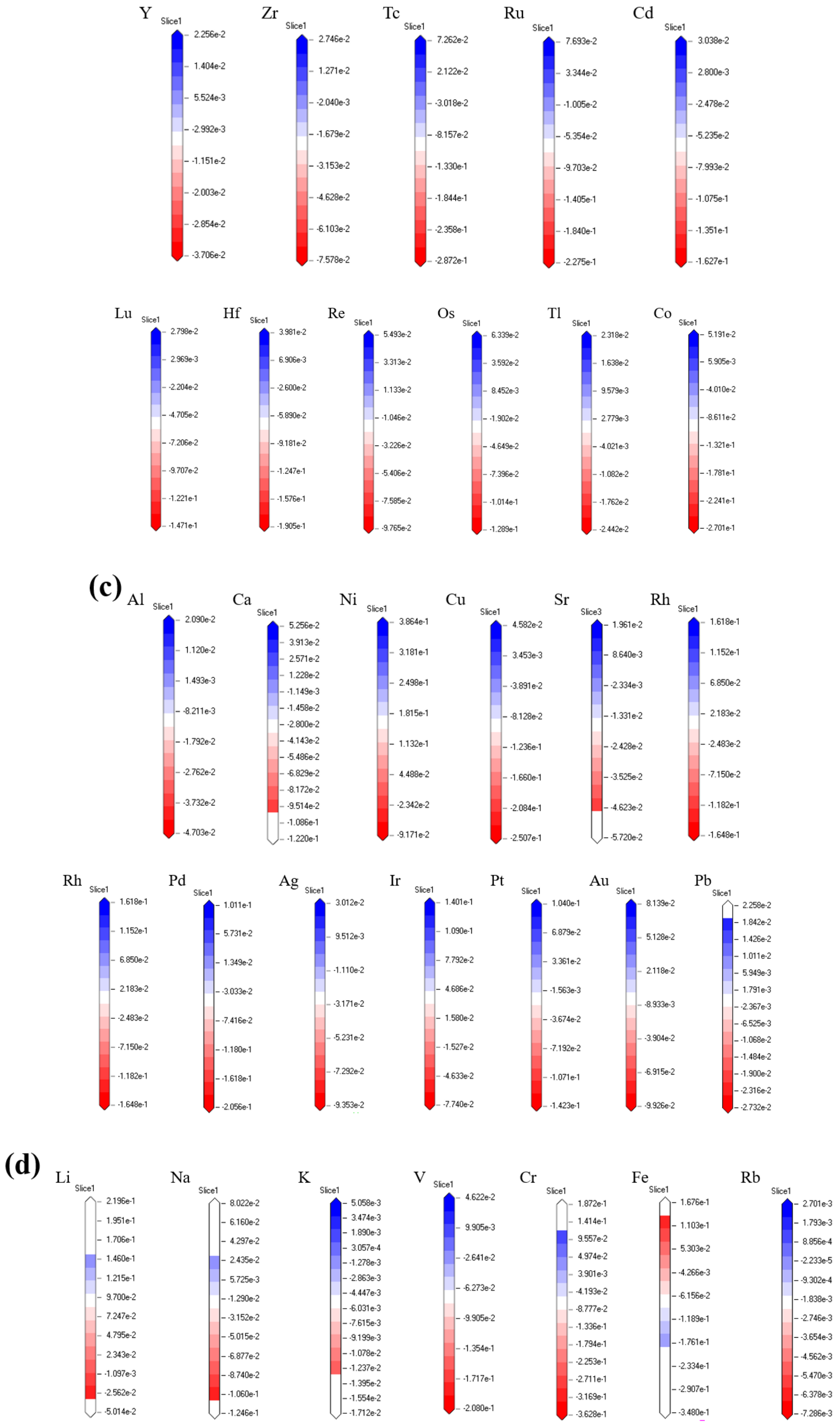

Y Slice1 2.256e-2 1.404e-2 5.524e-3 -2.992e-3 -1.151e-2 -2.003e-2 -2.854e-2 -3.706e-2
Zr Slice1 2.746e-2 1.271e-2 -2.040e-3 -1.679e-2 -3.153e-2 -4.628e-2 -6.103e-2 -7.578e-2
Tc Slice1 7.262e-2 2.122e-2 -3.018e-2 -8.157e-2 -1.330e-1 -1.844e-1 -2.358e-1 -2.872e-1
Ru Slice1 7.693e-2 3.344e-2 -1.005e-2 -5.354e-2 -9.703e-2 -1.405e-1 -1.840e-1 -2.275e-1
Cd Slice1 3.038e-2 2.800e-3 -2.478e-2 -5.235e-2 -7.993e-2 -1.075e-1 -1.351e-1 -1.627e-1
Lu Slice1 2.798e-2 2.969e-3 -2.204e-2 -4.705e-2 -7.206e-2 -9.707e-2 -1.221e-1 -1.471e-1
Hf Slice1 3.981e-2 6.906e-3 -2.600e-2 -5.890e-2 -9.181e-2 -1.247e-1 -1.576e-1 -1.905e-1
Re Slice1 5.493e-2 3.313e-2 1.133e-2 -1.046e-2 -3.226e-2 -5.406e-2 -7.585e-2 -9.765e-2
Os Slice1 6.339e-2 3.592e-2 8.452e-3 -1.902e-2 -4.649e-2 -7.396e-2 -1.014e-1 -1.289e-1
Tl Slice1 2.318e-2 1.638e-2 9.579e-3 2.779e-3 -4.021e-3 -1.082e-2 -1.762e-2 -2.442e-2
Co Slice1 5.191e-2 5.905e-3 -4.010e-2 -8.611e-2 -1.321e-1 -1.781e-1 -2.241e-1 -2.701e-1
(c)
Al Slice1 2.090e-2 1.120e-2 1.493e-3 -8.211e-3 -1.792e-2 -2.762e-2 -3.732e-2 -4.703e-2
Ca Slice1 5.256e-2 3.913e-2 2.571e-2 1.228e-2 -1.149e-3 -1.458e-2 -2.800e-2 -4.143e-2 -5.486e-2 -6.829e-2 -8.172e-2 -9.514e-2 -1.086e-1 -1.220e-1
Ni Slice1 3.864e-1 3.181e-1 2.498e-1 1.815e-1 1.132e-1 4.488e-2 -2.342e-2 -9.171e-2
Cu Slice1 4.582e-2 3.453e-3 -3.891e-2 -8.128e-2 -1.236e-1 -1.660e-1 -2.084e-1 -2.507e-1
Sr Slice3 1.961e-2 8.640e-3 -2.334e-3 -1.331e-2 -2.428e-2 -3.525e-2 -4.623e-2 -5.720e-2
Rh Slice1 1.618e-1 1.152e-1 6.850e-2 2.183e-2 -2.483e-2 -7.150e-2 -1.182e-1 -1.648e-1
Rh Slice1 1.618e-1 1.152e-1 6.850e-2 2.183e-2 -2.483e-2 -7.150e-2 -1.182e-1 -1.648e-1
Pd Slice1 1.011e-1 5.731e-2 1.349e-2 -3.033e-2 -7.416e-2 -1.180e-1 -1.618e-1 -2.056e-1
Ag Slice1 3.012e-2 9.512e-3 -1.110e-2 -3.171e-2 -5.231e-2 -7.292e-2 -9.353e-2
Ir Slice1 1.401e-1 1.090e-1 7.792e-2 4.686e-2 1.580e-2 -1.527e-2 -4.633e-2 -7.740e-2
Pt Slice1 1.040e-1 6.879e-2 3.361e-2 -1.563e-3 -3.674e-2 -7.192e-2 -1.071e-1 -1.423e-1
Au Slice1 8.139e-2 5.128e-2 2.118e-2 -8.933e-3 -3.904e-2 -6.915e-2 -9.926e-2
Pb Slice1 2.258e-2 1.842e-2 1.426e-2 1.011e-2 5.949e-3 1.791e-3 -2.367e-3 -6.525e-3 -1.068e-2 -1.484e-2 -1.900e-2 -2.316e-2 -2.732e-2
(d)
Li Slice1 2.196e-1 1.951e-1 1.706e-1 1.460e-1 1.215e-1 9.700e-2 7.247e-2 4.795e-2 2.343e-2 -1.097e-3 -2.562e-2 -5.014e-2
Na Slice1 8.022e-2 6.160e-2 4.297e-2 2.435e-2 5.725e-3 -1.290e-2 -3.152e-2 -5.015e-2 -6.877e-2 -8.740e-2 -1.060e-1 -1.246e-1
K Slice1 5.058e-3 3.474e-3 1.890e-3 3.057e-4 -1.278e-3 -2.863e-3 -4.447e-3 -6.031e-3 -7.615e-3 -9.199e-3 -1.078e-2 -1.237e-2 -1.395e-2 -1.554e-2 -1.712e-2
V Slice1 4.622e-2 9.905e-3 -2.641e-2 -6.273e-2 -9.905e-2 -1.354e-1 -1.717e-1 -2.080e-1
Cr Slice1 1.872e-1 1.414e-1 9.557e-2 4.974e-2 3.901e-3 -4.193e-2 -8.777e-2 -1.336e-1 -1.794e-1 -2.253e-1 -2.711e-1 -3.169e-1 -3.628e-1
Fe Slice1 1.676e-1 1.103e-1 5.303e-2 -4.266e-3 -6.156e-2 -1.189e-1 -1.761e-1 -2.334e-1 -2.907e-1 -3.480e-1
Rb Slice1 2.701e-3 1.793e-3 8.856e-4 -2.233e-5 -9.302e-4 -1.838e-3 -2.746e-3 -3.654e-3 -4.562e-3 -5.470e-3 -6.378e-3 -7.286e-3

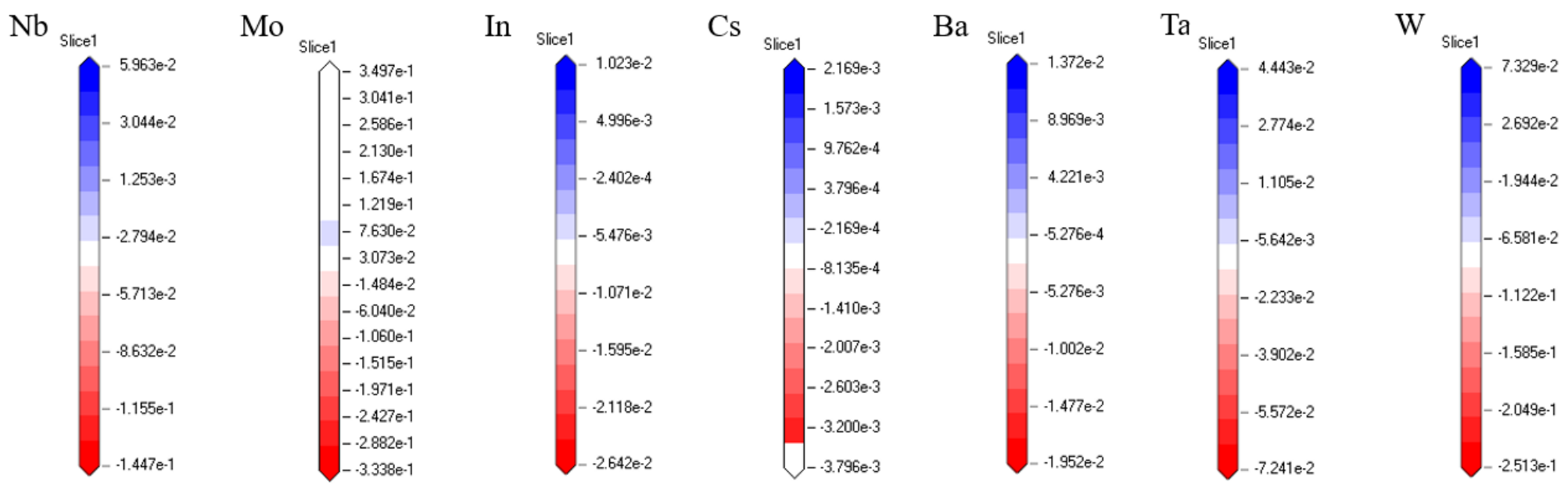

**Fig. S5** The value of differential charge density diagram.

**Table S1** Lattice parameters of 46 metal and semiconductor elements.

| Element | Angle | | | Lattice parameters | | |
|---|---|---|---|---|---|---|
| | α | β | γ | a (Å) | b (Å) | c (Å) |
| Li | 90° | 90° | 90° | 3.509 | 3.509 | 3.509 |
| Na | 90° | 90° | 90° | 4.291 | 4.291 | 4.291 |
| K | 90° | 90° | 90° | 5.320 | 5.320 | 5.320 |
| Rb | 90° | 90° | 90° | 5.700 | 5.700 | 5.700 |
| Cs | 90° | 90° | 90° | 6.140 | 6.140 | 6.140 |
| Be | 90° | 90° | 120° | 2.286 | 2.286 | 3.583 |
| Mg | 90° | 90° | 120° | 3.209 | 3.209 | 5.211 |
| Ca | 90° | 90° | 90° | 5.582 | 5.582 | 5.582 |
| Sr | 90° | 90° | 90° | 6.085 | 6.085 | 6.085 |
| Ba | 90° | 90° | 90° | 5.019 | 5.019 | 5.019 |
| Sc | 90° | 90° | 120° | 3.308 | 3.308 | 5.265 |
| Y | 90° | 90° | 120° | 3.645 | 3.645 | 5.731 |
| Lu | 90° | 90° | 120° | 3.505 | 3.505 | 5.549 |
| Ti | 90° | 90° | 120° | 2.951 | 2.951 | 4.679 |
| Zr | 90° | 90° | 120° | 3.231 | 3.231 | 3.231 |
| Hf | 90° | 90° | 120° | 3.195 | 3.195 | 5.051 |
| V | 90° | 90° | 90° | 3.028 | 3.028 | 3.028 |
| Nb | 90° | 90° | 90° | 3.301 | 3.301 | 3.301 |
| Ta | 90° | 90° | 90° | 3.303 | 3.303 | 3.303 |
| Cr | 90° | 90° | 90° | 2.885 | 2.885 | 2.885 |
| Mo | 90° | 90° | 90° | 3.147 | 3.147 | 3.147 |
| W | 90° | 90° | 90° | 3.165 | 3.165 | 3.165 |
| Tc | 90° | 90° | 120° | 2.735 | 2.735 | 4.388 |
| Re | 90° | 90° | 120° | 2.760 | 2.760 | 4.458 |
| Fe | 90° | 90° | 90° | 2.866 | 2.866 | 2.866 |
| Ru | 90° | 90° | 120° | 2.706 | 2.706 | 4.282 |
| Os | 90° | 90° | 120° | 2.735 | 2.735 | 2.735 |
| Co | 90° | 90° | 90° | 3.544 | 3.544 | 3.544 |
| Rh | 90° | 90° | 90° | 3.804 | 3.804 | 3.804 |
| Ir | 90° | 90° | 90° | 3.839 | 3.839 | 3.839 |
| Ni | 90° | 90° | 90° | 3.524 | 3.524 | 3.524 |
| Pd | 90° | 90° | 90° | 3.891 | 3.891 | 3.891 |
| Pt | 90° | 90° | 90° | 3.924 | 3.924 | 3.924 |
| Cu | 90° | 90° | 90° | 3.615 | 3.615 | 3.615 |
| Ag | 90° | 90° | 90° | 4.086 | 4.086 | 4.086 |
| Au | 90° | 90° | 90° | 4.078 | 4.078 | 4.078 |
| Zn | 90° | 90° | 120° | 2.665 | 2.665 | 4.947 |
| Cd | 90° | 90° | 120° | 2.979 | 2.979 | 5.617 |
| Al | 90° | 90° | 90° | 4.050 | 4.050 | 4.050 |
| In | 90° | 90° | 90° | 3.252 | 3.252 | 4.951 |
| Tl | 90° | 90° | 120° | 4.457 | 4.457 | 5.525 |
| C | 90° | 90° | 90° | 3.567 | 3.567 | 3.567 |
| Si | 90° | 90° | 90° | 5.431 | 5.431 | 5.431 |
| Ge | 90° | 90° | 90° | 5.658 | 5.658 | 5.658 |
| Sn | 90° | 90° | 90° | 6.491 | 6.491 | 6.491 |
| Pb | 90° | 90° | 90° | 4.950 | 4.950 | 4.950 |

**Table S2** Cut-off energies and *k*-points of 46 metal and semiconductor elements.

| Element | Cut-off energy | *k*-point |
|---|---|---|
| Li | 400.0 eV | 8×8×8 |
| Na | 400.0 eV | 6×6×6 |
| K | 435.4 eV | 6×6×6 |
| Rb | 400.0 eV | 4×4×4 |
| Cs | 272.1 eV | 4×4×4 |
| Be | 571.4 eV | 13×13×8 |
| Mg | 479.8 eV | 9×9×6 |
| Ca | 381.0 eV | 4×4×4 |
| Sr | 250.3 eV | 4×4×4 |
| Ba | 190.5 eV | 6×6×6 |
| Sc | 410.9 eV | 9×9×6 |
| Y | 272.1 eV | 8×8×4 |
| Lu | 408.2 eV | 8×8×6 |
| Ti | 381.0 eV | 10×10×6 |
| Zr | 272.1 eV | 9×9×6 |
| Hf | 435.4 eV | 9×9×6 |
| V | 381.0 eV | 8×8×8 |
| Nb | 400.0 eV | 8×8×8 |
| Ta | 359.2 eV | 8×8×8 |
| Cr | 1197.0 eV | 10×10×10 |
| Mo | 400.0 eV | 8×8×8 |
| W | 353.7 eV | 8×8×8 |
| Tc | 419.1 eV | 11×11×6 |
| Re | 347.3 eV | 10×10×6 |
| Fe | 1306.0 eV | 10×10×10 |
| Ru | 449.0 eV | 11×11×6 |
| Os | 370.1 eV | 11×11×6 |
| Co | 381.0 eV | 8×8×8 |
| Rh | 549.7 eV | 8×8×8 |
| Ir | 326.5 eV | 8×8×8 |
| Ni | 400.0 eV | 8×8×8 |
| Pd | 560.6 eV | 6×6×6 |
| Pt | 321.1 eV | 6×6×6 |
| Cu | 408.2 eV | 8×8×8 |
| Ag | 517.0 eV | 6×6×6 |
| Au | 321.1 eV | 6×6×6 |
| Zn | 394.6 eV | 11×11×6 |
| Cd | 299.3 eV | 10×10×4 |
| Al | 190.5 eV | 6×6×6 |
| In | 326.5 eV | 8×8×5 |
| Tl | 340.1 eV | 8×8×6 |
| C | 400.0 eV | 8×8×8 |
| Si | 190.5 eV | 6×6×6 |
| Ge | 394.6 eV | 4×4×4 |
| Sn | 312.9 eV | 4×4×4 |
| Pb | 400.0 eV | 6×6×6 |

**Table S3** LDA, PBE and HSE06 functionals were used to calculate the band gap values (eV) and experimental band gap values (eV) of C, Si and Ge.

| EI | LDA | PBE | HSE06 | EXP |
|---|---|---|---|---|
| C | 3.934 | 3.988 | 5.039 | 5.45 |
| Si | 0.512 | 0.591 | 1.152 | 1.12 |
| Ge | 0.449 | 0.494 | 0.968 | 0.67 |